\documentclass[nofootinbib,a4paper,notitlepage]{revtex4-1}

\usepackage{graphicx,psfrag}
\usepackage{mathrsfs}
\usepackage{amsmath,amsfonts,amssymb}
\usepackage{multirow}
\usepackage{comment}
\usepackage{hyperref}

\usepackage{xspace}
\usepackage[labelfont=bf,list=true]{subcaption}
\usepackage[labelfont=bf]{caption}

\usepackage{mathrsfs}
\DeclareSymbolFontAlphabet{\mathrsfs}{rsfs}
\newcommand{\scri}{\mathrsfs{I}}
\newcommand{\scripx}{$\scri^+$}
\newcommand{\scrip}{\scripx\xspace}

\newcommand{\thesis}{Vano-Vinuales:2015lhj}
\newcommand{\procere}{Vano-Vinuales:2014ada}
\newcommand{\procmg}{Vano-Vinuales:2016mbo}
\newcommand{\pap}{Vano-Vinuales:2014koa}
\newcommand{\papgauge}{Vano-Vinuales:2017qij}

\newcommand{\rscri}{r_{\!\!\scri}}
\newcommand{\Kc}{K_{CMC}}
\newcommand{\Cc}{C_{CMC}}
\newcommand{\nconst}{n_{cK}}

\newcommand{\cnK}{\xi_{cK}}

\newcommand{\backbetar}{\hat{\beta^r}}
\newcommand{\backalpha}{\hat{\alpha}}

\newcommand{\cp}{c}
\newcommand{\const}{\cp^2}

\newcommand{\CZ}{Z4c}
\newcommand{\aconf}{\bar\Omega}

\newcommand{\K}{\Delta \tilde K}
\newcommand{\hpt}{h'(\tilde r)}
\newcommand{\cosm}{\kappa_0}

\newcommand{\statiox}{statio}
\newcommand{\statio}{\statiox\xspace}

\newcommand{\eref}[1]{(\ref{#1})}
\newcommand{\sref}[1]{section~\ref{#1}}
\newcommand{\ssref}[1]{subsection~\ref{#1}}
\newcommand{\aref}[1]{appendix~\ref{#1}}
\newcommand{\fref}[1]{figure~\ref{#1}}
\newcommand{\sfref}[1]{subfigure~\ref{#1}}

\begin{document}

\title{Spherically symmetric black hole spacetimes on hyperboloidal slices} 

\author{Alex Va\~n\'o-Vi\~nuales} 

\affiliation{CENTRA, Departamento de F\'isica, Instituto  Superior T\'ecnico IST, Universidade de Lisboa UL, Avenida Rovisco Pais 1, 1049-001 Lisboa, Portugal}

\begin{abstract}

Gravitational radiation and some global properties of spacetimes can only be unambiguously measured at future null infinity (\scrip). This motivates the interest in reaching it within simulations of coalescing compact objects, whose waveforms are extracted for gravitational wave modelling purposes. One promising method to include future null infinity in the numerical domain is the evolution on hyperboloidal slices: smooth spacelike slices that reach future null infinity. The main challenge in this approach is the treatment of the compactified asymptotic region at \scrip. 
Evolution on a hyperboloidal slice of a spacetime including a black hole entails an extra layer of difficulty, in part due to the finite coordinate distance between the black hole and future null infinity.  
Spherical symmetry is considered here as simplest setup still encompassing the full complication of the treatment along the radial coordinate.  
First, the construction of constant-mean-curvature hyperboloidal trumpet slices for Schwarzschild and Reissner-Nordstr\"om black hole spacetimes is reviewed from the point of view of the puncture approach. Then, the framework is set for solving hyperboloidal-adapted hyperbolic gauge conditions for stationary trumpet initial data, providing solutions for two specific sets of parameters. Finally, results of testing these initial data in evolution are presented. 

\end{abstract}

\maketitle

\section{Introduction}

The accurate numerical treatment of black holes (BHs) and their emitted gravitational wave (GW) signals is primordial for the field of GW astronomy. BHs are the most common participants in the compact binary coalescences observed so far \cite{LIGOScientific:2019lzm,LIGOScientific:2023vdi}, but are challenging to model numerically due to the presence of the physical singularity inside of their horizon. GWs, as radiation propagating at the speed of light, are only unambiguously defined at future null infinity \scrip, the collection of the end points of future-directed null geodesics. Future null infinity also corresponds to the idealized location of observers of astrophysical events \cite{Barack:1998bv,Leaver1986,PhysRevD.34.384}, such as GW interferometers, so that is where GWs signals should ideally be extracted from simulations. 

Two main description of BHs are common in numerically simulated spacetimes. Excision \cite{Seidel:1992vd} involves setting an artificial timelike inner boundary inside the BH horizon to avoid the slices from reaching the physical singularity. This exploits the fact that no physical information is allowed to exit the BH, but the need to know the location of the apparent horizon at all times makes this approach technically difficult for generic spacetimes. Still, it has been successfully used to produce the largest, longest and most accurate binary BH waveform catalog currently available \cite{Boyle:2019kee}. In the puncture method, a specific singularity-avoidant slice of the BH spacetime is considered. This slice can have the topology of a wormhole, where the asymptotically flat end at the other side of the BH is compactified and represents the BH's location \cite{1963PhRv..131..471B,Brandt:1997tf,Beig:1994rp}. In evolutions of wormhole puncture initial data with the ``moving puncture'' gauge \cite{Campanelli:2005dd,Baker:2005vv}, the initial slice numerically detaches from the asymptotically flat end beyond the horizon and its topology becomes close to that of a compactified trumpet \cite{Hannam:2006xw,Hannam:2006vv,Hannam:2008sg}, where the proper distance becomes infinite while reaching towards the symmetric point to future timelike infinity $i^+$. For embedding diagrams of the wormhole and trumpet geometries see e.g. figures 1 and 2 in \cite{Hannam:2008sg}. Construction of maximal trumpet slices has been tackled for Schwarzschild \cite{Baumgarte:2007ht,Dennison:2014sma,Bruegmann:2009hof,Baumgarte:2022auj,Li:2023pme}, for Reissner-Nordstr\"om (RN) \cite{Li:2022pev}, for Kerr \cite{Dennison:2014eta,Heissel:2017hxm}. Asymptotically, the slices considered in those works are spacelike Cauchy, and thus reach spatial infinity $i^0$. The trumpet puncture approach is also chosen in the present work, for its simpler technical implementation and for the possibility to reach a portion inside of the horizon. While the latter is not required for GW extraction, it can provide insights into the numerical behaviour of slices inside of the horizon, useful e.g. for the construction of Penrose diagrams of dynamical scenarios \cite{penrosepap}.

Including future null infinity within the numerical integration domain is possible by evolving on a suitable choice of foliation. The most straightforward option are characteristic slices, which can provide considerable simplifications in the equation used \cite{Winicour2009LRR}, but are prone to the development of caustics. 
Cauchy-characteristic matching \cite{BishopCCM,Szilagyi:2000xu,Winicour2009LRR} joins an inner Cauchy spacelike slice to an outer characteristic one along a timelike boundary. However, compatible formulations of the Einstein equations for each domain are required. 
In Cauchy-characteristic evolution \cite{Bishop:1996gt,Reisswig:2009us,Babiuc:2011qi,Moxon:2021gbv} the same setup is used, but the Cauchy evolution is performed independently and then used as inner boundary data for the characteristic evolution.
A more flexible and elegant alternative is the evolution on hyperboloidal \cite{friedrich1983,Friedrich86,lrr-2004-1,Friedrich:2003fq} slices, which are spacelike and reach null infinity. 
An advantage that hyperboloidal evolution is expected to have and that has been achieved with Cauchy-characteristic evolution is resolving GW memory \cite{Mitman:2020pbt}.
A radial compactification on hyperboloidal slices allows to include future null infinity in a finite domain. Unlike a compactification of Cauchy slices where radiation travelling out is slowed down and becomes underresolved, the outward propagation speed of signals on compactified hyperboloidal slices is of order unity and they reach \scrip at a finite coordinate time without any loss of resolution. Figure~5 in \cite{\pap} illustrates this effect with a scalar field perturbation.

Conformal compactification \cite{PhysRevLett.10.66} is one method to tackle compactified hyperboloidal slices, which allow us to reach \scrip with a finite value of the coordinates. The core idea is that instead of working with the physical metric $\tilde g_{ab}$ that diverges at infinity when the coordinates are compactified, the Einstein equations are instead expressed in terms of a finite conformally rescaled metric $\bar g_{ab}$, related to the physical one by a conformal factor $\Omega$ that vanishes at $\scri^+$ at the appropriate rate
\begin{equation}\label{rescmetric}
\bar g_{ab} = \Omega^2\tilde g_{ab} . 
\end{equation}
One of the most difficult aspects of this approach to the hyperboloidal initial value problem \cite{Frauendiener:1997ze,Friedrich:2003fq} is the regularisation of the resulting formally singular equations (see \eref{eq:EEconformal} in \sref{formulation}) in a way that works numerically and avoids instabilities arising from the continuum equations, in particular for hyperbolic free evolution schemes as considered here. 
At the analytical level, the equations were shown to be manifestly regular at \scrip \cite{friedrich1983,Friedrich:2003fq}, however that specific formulation does not treat BHs in a straightforward way and suffers from continuum instabilities \cite{Husa:2002kk}. In contrast to the conformal approach, the dual foliation method \cite{Hilditch:2015qea,Hilditch:2016xzh}, a generalization of the dual coordinate frame method used in \cite{Scheel:2006gg}, aims to minimise the divergent terms in the equations, making them as regular as possible. 
The present implementation follows Zengino\u{g}lu's approach \cite{Zenginoglu:2007jw,Zenginoglu:2008pw,Zenginoglu:2008wc,Zenginoglu:2007it} to conformal compactification, using free evolution and a time-independent conformal factor $\Omega$. Stable evolutions in spherical symmetry of regular initial data that do not form BHs were presented in \cite{\pap}, while \cite{\papgauge} covers experiments with suitable hyperbolic gauge conditions. 

Evolving a hyperboloidal slice of spacetime including BHs is particularly challenging\footnote{Past experiments in spherical symmetry (subsection 8.2.1 in \cite{\thesis}, also mentioned here at the end of \ssref{ss:gaugebh}) have shown an instability-inducing drift in the variables, not linked to any specific part of the domain. This was related mainly to gauge conditions and how they deal with the trumpet and \scrip asymptotics.}, especially in the puncture approach where both the regions inside of the horizons and the asymptotic far field are compactified.
Constant-mean-curvature (CMC) foliations, where the trace of the physical extrinsic curvature takes a constant value, are well known in the literature, e.g. for the Schwarzschild \cite{Malec:2009hg,Cruz-Osorio:2010rsr,Lee:2011dp} and RN \cite{Tuite:2013hza,Lee:2018kbj} spacetimes. Of special interest are those specific CMC slices that correspond to trumpet slices in their corresponding BH geometry: in a certain way these are a generalization of the maximal trumpet slices mentioned above. The difference is that CMC slices with non-vanishing trace of the extrinsic curvature asymptotically reach null infinity, and thus can be used as hyperboloidal trumpet slices suitable for evolving a BH spacetime all the way to future null infinity. 

Several works have considered hyperboloidal initial data including BHs. Configurations in spherical and in axial symmetry were presented in \cite{Schneemann}, while \cite{Schinkel:2013tka} considered axisymmetric CMC slices for Kerr and \cite{Schinkel:2013zm} studied perturbed Kerr initial data on asymptotic CMC slices. The generalization of Bowen-York initial data to hyperboloidal slices for binaries of boosted and spinning BHs was carried out in \cite{Buchman:2009ew}, whereas properties such as the Bondi-Sachs energy and momentum of the above setups were presented in \cite{Bardeen:2011pd}. The binary BH scenario was also studied in \cite{schinkelthesis}. However, these works were designed with the aim to treat the BHS via excision, and thus not a lot of effort was put into regularising the slices beyond the BH horizon. 

The description of BHs via punctures requires a careful treatment of the hyperboloidal slices inside the BHs as well. In previous work \cite{\procere,\procmg,\thesis}, evolution of hyperboloidal CMC Schwarzschild trumpet initial data was considered, as well as the collapse into a BH of a scalar field perturbation on a regular spacetime. The trumpet dynamics was found to be highly dependent on the choice of gauge conditions. CMC trumpet initial data stationary with respect to the given gauge conditions are very desired, as the evolution of any perturbation on these initial data would be easier to identify and study. Imposing stationarity is the approach suggested in \cite{Ohme:2009gn}, although the slicing condition considered there is most likely not appropriate for numerical evolutions. In previous numerical experiments, a stationary solution is reached by the evolution at late times (such as that on the right of figure 2 in \cite{\procmg}), be it with BH trumpet or collapsing scalar field initial data, for at least some choices of gauge conditions. It thus makes sense to consider stationary solutions of the gauge conditions as candidates for an initial hyperboloidal trumpet slice. 

The aims of this work are to review hyperbolic CMC trumpet BH initial data suitable for numerical evolutions with the puncture approach in mind (\sref{cmcsec}), and to set the basic infrastructure in terms of initial data and gauge conditions to calculate stationary trumpet slices (\sref{statio}). 
An example of such a stationary configuration is solved for a specific choice of gauge, and basic evolutions for both CMC and solved-for initial data are performed on hyperboloidal slices (\sref{evolsec}). For this purpose, the already nontrivial hyperboloidal initial value problem in spherical symmetry is considered, as it still contains the critical part of the radial treatment. 

This paper is organised as follows: in \sref{formulation} the used formulation of the conformally compactified Einstein equations is briefly described, and the gauge conditions considered here are covered in \sref{gauge}. Initial data including a BH is treated in the following two sections: as constant-mean-curvature (CMC) in \sref{cmcsec}, while an example of solving hyperboloidal-adapted gauge conditions is provided in \sref{statio}. Section~\ref{evolsec} presents basic evolution results, and final thoughts on this work are gathered in the conclusions. The appendix collects an equation used in \sref{statio}. Sections \ref{formulation}, \ref{gauge} and \ref{cmcsec} cover previously treated material, while sections \ref{statio} and \ref{evolsec} present new research. 

The chosen metric signature is $(-,+,+,+)$ and, as is customary, the fundamental constants are set to $G=c=1$. The convention for the sign of the extrinsic curvature is that of Misner, Thorne, and Wheeler \cite{Misner1973}, meaning that a negative\footnote{This is why the constant parameter $\Kc$ introduced in \eref{ein:omega} and described in \ssref{ss:htf} is negative for hyperboloidal slices reaching future null infinity. If a positive value is chosen for it, then the hyperboloidal slices intersect past null infinity.} value means expansion of the normals. Notation for the metrics is the same one as used in \cite{\pap}: the 4-dimensional physical spacetime metric is denoted as $\tilde g$, the 4D conformal metric as $\bar g$, the 3D conformal spatial metric (induced by $\bar g$) as $\bar \gamma$, the 3D twice conformal metric $\gamma$ and the 3D twice conformal background metric $\hat \gamma$.

\section{Formulation} \label{formulation}

The emphasis in this work is on hyperboloidal BH initial data, so only a brief review of the formulation of the evolved system with corresponding references is given. 
Expressed in terms of the rescaled metric $\bar g_{ab}$ as defined in \eref{rescmetric}, the 4D Einstein equations take the form
\begin{equation}\label{eq:EEconformal}
G_{ab}[\bar g] = 8\pi\ T_{ab} -\frac{2}{\Omega}\left(\bar \nabla_a\bar \nabla_b\Omega-\bar g_{ab}\bar \nabla^c\bar \nabla_c\Omega\right)-\frac{3}{\Omega^2}\bar g_{ab}(\bar \nabla_c\Omega)\bar \nabla^c\Omega . 
\end{equation}
The well-posed formulations considered are either the Generalized BSSN \cite{NOK,PhysRevD.52.5428,Baumgarte:1998te,Brown:2007nt} or a similar conformal version of the Z4 \cite{bona-2003-67,Alic:2011gg,Sanchis-Gual:2014nha}, the \CZ{} equations \cite{Bernuzzi:2009ex,Weyhausen:2011cg}. The full derivation of these equations in terms of the conformally rescaled metric is described in \cite{\pap} and in Chapter 2 of \cite{\thesis}. The equations used in the simulations are those included in appendix C of \cite{\pap} (or appendix A in \cite{\papgauge}) and again in Chapter 2 of \cite{\thesis}. There is a modification related to the evolution of BH spacetimes: a constraint damping term of the form 
\begin{equation} \label{e:extradamping} -\cosm\frac{Z_r}{r}, \end{equation}
where $Z_r$ is a Z4 variable and $\cosm$ a freely specifiable parameter, can be added to $\dot\Lambda^r$'s right-hand-side (RHS). This term helps suppress instabilities if extrapolating boundary conditions are used at $r=0$ (this was not necessary for evolutions of regular spacetimes, as parity conditions could be imposed at the origin).

The evolution variables are the 3D conformally rescaled spatial metric 
\begin{equation}
\gamma_{ab}=\chi\bar \gamma_{ab} , 
\end{equation}
where $\bar \gamma_{ab}$ is the spatial metric induced from $\bar g_{ab}$, and $\chi$ is the spatial conformal factor.
The conformal extrinsic curvature tensor $\bar K_{ab}$ is decomposed into its conformal trace-free part 
\begin{equation}
A_{ab}=\chi\bar K_{ab}-\frac{1}{3}\gamma_{ab}\bar K ,  \quad \textrm{with} \quad \bar K=\bar K_{ab}\bar\gamma^{ab}\equiv K_{ab}\gamma^{ab}, 
\end{equation}
 and (in this formulation) its physical trace, mixed with the physical Z4 variable $\tilde \Theta$, 
\begin{equation}
\tilde K = \Omega\bar K-\frac{3\beta^a\partial_a\Omega}{\alpha}-2\tilde\Theta. , 
\end{equation} 
Evolved are $A_{ab}$, and $\tilde K$'s variation with respect to its initial value $\Delta\tilde{K}=\tilde K-\tilde K_0=\tilde K-\Kc$ (this last parameter will be explained in \ssref{ss:htf}). The quantity $\tilde \Theta$ is evolved as well if using the Z4 formulation.
The Z4 variable $Z_a$ is absorbed into the vector 
\begin{equation}\label{Lambdadef}
\Lambda^a=\gamma^{bc}\left(\Gamma^a_{bc}-\hat\Gamma^a_{bc}\right) +2\gamma^{ab}Z_b ,
\end{equation} 
where $\Gamma^a_{bc}$ are the Christoffel symbols calculated from $\gamma_{ab}$ and $\hat\Gamma^a_{bc}$ the ones built from a time-independent background metric $\hat \gamma_{ab}$. The latter is chosen to be the flat spatial metric in spherical coordinates, and its explicit components (following an equivalent notation to that in \eref{e:linel} are given in \eref{e:hatvals}.
The evolved gauge variables are the conformal lapse $\alpha$ and the shift $\beta^i$.

\subsection{Spherically symmetric reduction variables}\label{sphersym}

The following spherically symmetric ansatz is used for the spherically symmetric line element in the conformally compactified domain (with $\sigma^2\equiv d\theta^2+\sin^2\theta d\phi^2$)
\begin{equation}\label{e:linel}
ds^2 = - \left(\alpha^2-\chi^{-1}\gamma_{rr}{\beta^r}^2\right) dt^2 + \chi^{-1}\left[2\, \gamma_{rr}\beta^r dt\,dr +  \gamma_{rr}\, dr^2 +  \gamma_{\theta\theta}\, r^2\, d\sigma^2\right] ,
\end{equation}
where the freedom introduced by the spatial conformal factor $\chi$ is fixed by eliminating $\gamma_{\theta\theta}=\gamma_{rr}^{-1/2}$. 

In spherical symmetry, the only independent component of the trace-free part of the conformal extrinsic curvature $A_{ab}$ after explicitly imposing its trace-freeness is $A_{rr}$. Also only the radial component of the quantities $\Lambda^a$, $\beta^a$ and $Z_a$ (denoted by $\Lambda^r$, $\beta^r$ and $Z_r$ respectively) remains non-zero. The evolution variables of the spherically symmetric reduced system are $\chi$, $\gamma_{rr}$, $A_{rr}$, $\K$, $\Lambda^r$, $\alpha$, $\beta^r$ and $\tilde\Theta$.

The conformal factor $\Omega$ is set to be a time-independent function of the compactified radial coordinate $r$ as 
\begin{equation}\label{ein:omega}
\Omega(r) = \left(-\Kc\right)\frac{\rscri^2-r^2}{6\, \rscri},
\end{equation}
with $\rscri$ the coordinate location of future null infinity (set to $\rscri=1$ in the implementation without restricting generality) and $\Kc$ a negative parameter described in \ssref{ss:htf}. This expression satisfies that $\Omega(r)$ is a regular function that becomes zero at $\scri$, with non-vanishing derivative there (compare e.g. \cite{Husa:2002zc,Schneemann}). The origin of this expression is explained in \ssref{ss:compact}. 

\section{Gauge conditions} \label{gauge}

Hyperboloidal constrained evolutions \cite{Moncrief:2008ie,Rinne:2009qx,Morales:2016rgt} have used suitable gauges imposed via the resolution of elliptic constraint equations. In this work the free evolution approach is employed for its faster performance in simulations, and it requires the use of hyperbolic gauge conditions. 
The gauge quantities, lapse $\alpha$ and shift $\beta^i$, control the behaviour of the coordinates, and they are critically important for a successful and efficient evolution. Bad choices will easily lead simulations to crash at an earlier or later time. An example of the effects of gauge choices in this hyperboloidal work is that they can induce deformations in propagating signals, as is illustrated by the (deformed) scalar field signals at \scrip in figure 2 in \cite{\papgauge}, that are to be corrected in post-processing.  

For vanishing cosmological constant and vacuum or compact support matter sources, future null infinity is an ingoing null hypersurface. This means that no information is allowed to enter the domain from the outside, making it a natural boundary for the numerical integration domain, where no boundary conditions need to be imposed -- radiation just needs to be allowed to leave the spacetime. It is possible to fix $\scri^+$ to a specific coordinate location in the numerical grid ($\rscri$ \eref{ein:omega} in the present setup) for compactified hyperboloidal slices.
This procedure is called scri-fixing \cite{Frauendiener:1997ze,Zenginoglu:2007jw}.  

The background behaviour of hyperboloidal slices differs from Cauchy ones in that the trace of the physical extrinsic curvature $\tilde K$ is non-zero asymptotically. This requires a modification of the usual slicing conditions commonly used in numerical simulations. See e.g. the generalizations of the Bona-Mass\'o family of slicing conditions \cite{Bona:1994dr} and the modifications of the Gamma-driver shift \cite{Alcubierre:2002kk} and harmonic shift conditions \cite{Friedrich:2000qv} included in \cite{\papgauge}. 
The basic idea behind those modifications is the addition of specific non-principal-part source terms to the gauge evolution equations, to ensure that a hyperboloidal slice of Minkowski spacetime will be a stationary solution of the gauge equations. 
This is described in the next subsection. 

An optimal prescription for hyperbolic gauge conditions for the conformally compactified hyperboloidal approach is still to be found. Experimentation with possible gauge source functions has provided several successful working examples. They are being further studied and extended, here by including a BH in the spacetime, and elsewhere by being tested in the full 3D case \cite{3dpaper}. Work towards finding suitable gauge conditions \cite{Duarte:2022vxn} is also being tackled from the dual foliation approach. 

\subsection{Hyperbolic gauge conditions tested with BH spacetimes}\label{ss:gaugebh}

\begin{figure}[htbp!!]
\center
\includegraphics[width=0.8\linewidth]{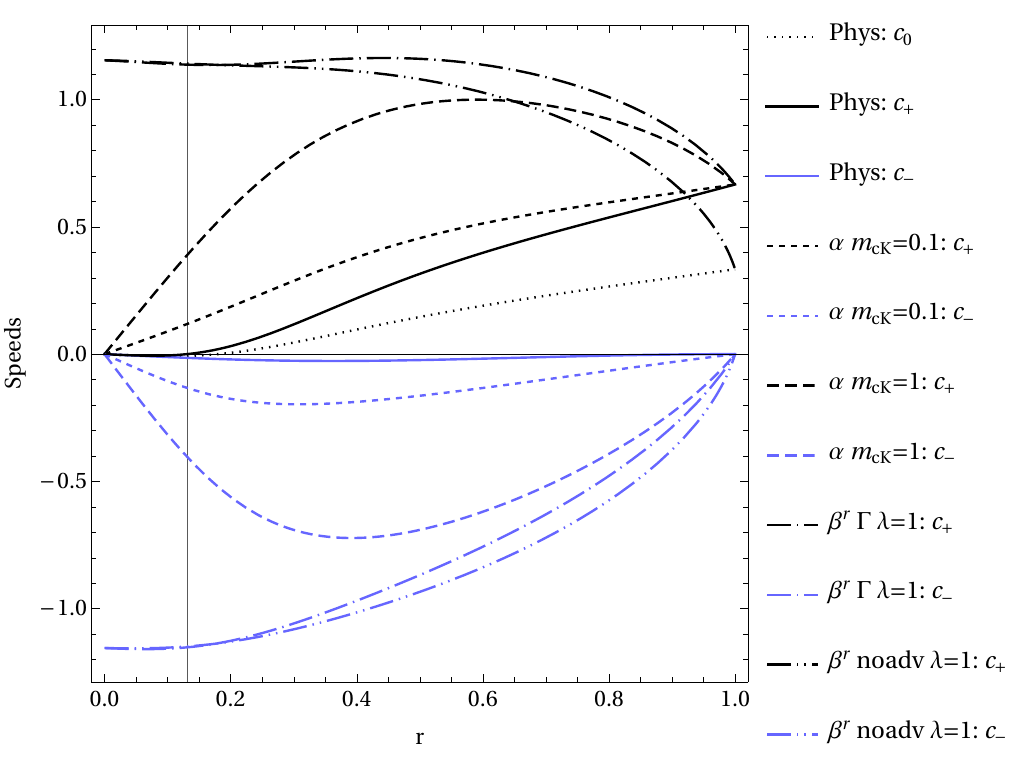}
\caption{Zero speed $c_0=-\beta^r$ and incoming ($-$) and outgoing ($+$) lightspeeds $c_\pm=-\beta^r\pm\alpha\sqrt{\frac{\chi}{\gamma_{rr}}}$ plotted for a CMC slice of a Schwarzschild BH with $M=1$, using $\Kc=-1$ and $\rscri=1$. The vertical line indicates the radial position of the horizon, where $c_0,c_-<0$ and $c_+=0$. At $\scri^+$ we have $c_0,c_+>0$ and $c_-=0$. Except for $c_0$, outgoing speeds are shown in black and ingoing ones in blue. The two sets of dashed lines correspond to the incoming and outgoing modified propagation speeds related to the slicing condition as in \eref{adotsolve} with the choices of parameter $m_{cK}$ used in \ssref{statiosolve}. While the speeds go to zero at the location of the trumpet ($r=0$) and coincide whith the lightspeeds at \scrip, their values are different in the rest of the domain. The two sets of dash-dot curves show the characteristic speeds associated with the shift condition with $\lambda=1$, for shift choices \eref{eg:integGammadriver} and \eref{modifiedshift}. In the second case, where the shift advection terms have been dropped, the incoming speed is made to be zero at \scrip, but that forced the outgoing one to be equal to $c_0$ (instead of $c_+$) there. Compare to a Minkowski equivalent (with different parameter choices) in figure 1 in \cite{\papgauge}.}
\label{fr:lightspeeds}
\end{figure}

When applying the gauge conditions discussed in \cite{\papgauge} to a BH spacetime, one important aspect is to recognize that harmonic slicing is only marginally singularity avoiding, which means that a singularity is reached in an infinite coordinate time. Harmonic slicing is thus not a good choice in the neighbourhood of a BH if excision is not used. However, near \scrip the physical propagation speeds of harmonic lapse (and shift) ensure that no unknown gauge information enters the numerical domain through future null infinity. Thus the optimal scenario is to use harmonic slicing near \scrip and something different close to the BH. 
A condition that has provided successful evolutions using trumpet initial data is, with $\dot{} \equiv \partial_t$ and $' \equiv \partial_r$ , 
\begin{equation}\label{adotsolve} 
\dot\alpha = \beta^r\alpha' -\backbetar\backalpha'-\frac{(\nconst+\alpha^2)\K}{\Omega}+\frac{\Omega'}{\Omega}(\backbetar\backalpha-\beta^r\alpha) + \frac{\cnK(\backalpha-\alpha)}{\Omega} , 
\end{equation}
{where $\cnK$ is a parameter used to damp the behaviour of the lapse at \scrip.} This equation is equivalent to (20) in \cite{\papgauge} with $\xi_1\hat\alpha=\cnK$, $\xi_2=0$ and $\alpha^2f= \nconst+\alpha^2$, later setting $\nconst$ \footnote{The time-independent quantity $\nconst$, a function of the radial coordinate, has here a different expression from that used in \cite{\papgauge}.} to be proportional to $\Omega$. 
Note that the coefficient in front of $\K$ is similar to the shock-avoiding slicing condition \cite{Alcubierre:1996su,Baumgarte:2022ecu,Li:2023pme}. 
This form was chosen for the following considerations. The $\alpha^2$ part provides physical propagation speeds for the gauge modes (the first three lines listed in \fref{fr:lightspeeds}), as mentioned above. This is desired at \scrip, because then all propagation speeds are either positive or zero, and there are no incoming modes there. However, near the location of the trumpet inside of the BH's horizon, the physical propagation speeds become zero (as $\alpha=0$ at the location of the trumpet). The effect is that any signals that have entered the BH region and travel along the infinitely long cylinder of the trumpet slice will propagate slower and slower, soon becoming underresolved, which can lead to numerical instabilities. Increasing the gauge propagation speeds allows perturbations to leave the domain in a finite time and provides more stable evolutions in general, and also gives smoother stationary values for the evolution quantities at the trumpet. Examples of modified propagation speeds for the lapse and shift conditions are shown in \fref{fr:lightspeeds}. 

These gauge source functions are designed to make a hyperboloidal CMC slice of Minkowski (encoded in the hatted quantities) a stationary solution of the slicing equation: $\dot\alpha=0\leftrightarrow\alpha=\backalpha$, $\beta^r=\backbetar$, $\K=0$. 
The components of the background conformally compactified metric (following an ansatz like that of \eref{e:linel}) that appear in \eref{adotsolve}, and are used to calculate $\hat \Gamma^a_{bc}$ in \eref{Lambdadef}, are
\begin{equation}\label{e:hatvals}
\hat\chi = \hat{\gamma_{rr}} = \hat{\gamma_{\theta\theta}} = 1 , \quad \hat\alpha = \sqrt{\Omega^2+\left(\frac{\Kc\,r}{3}\right)^2} \quad \textrm{and} \quad \hat\beta^r = \frac{\Kc\,r}{3}. 
\end{equation}
They are obtained (\ssref{ss:htf}) from \eref{varshypcomp} or \eref{varshypcompc} setting $A(\frac{r}{\aconf})=1$, $M=0$, $\Cc=0$ and $\aconf=\Omega$. 

For the shift condition, two different options are considered. One is a variant of the integrated Gamma-driver \cite{Alcubierre:2002kk} adapted to hyperboloidal slices 
\begin{equation}\label{eg:integGammadriver}
\dot{{\beta^r}} = {\beta^r} {\beta^r}'-\backbetar \backbetar'+\left(\lambda (\rscri^2-r^2) +\frac{3}{4}\alpha^2\chi\right) \Lambda^r+\eta  (\backbetar-{\beta^r})+ \xi_{\beta^r} \left(\frac{\backbetar}{\Omega }-\frac{{\beta^r}}{\Omega }\right) ,
\end{equation}
mostly the same as (26) in \cite{\papgauge}. The coefficient in front of $\Lambda^r$ is chosen in such a way that the associated propagation speeds will be the physical ones at \scrip. The positive parameter $\lambda$ will only increase the speeds near the trumpet, in a similar fashion as $\nconst$ for the slicing condition above. This is shown in \fref{fr:lightspeeds}. 
The other shift option is to have an expression purely proportional to $\Lambda^r$: the resulting system is still hyperbolic and it will have conformally flat initial data as a stationary solution (more on this in \sref{statio}). However, dropping the advection terms modifies the characteristic propagation associated to the shift condition. In order to ensure that the related ingoing speed at \scrip is still zero, the coefficient in front of $\Lambda^r$ is modified as 
\begin{equation}\label{modifiedshift}
\dot{{\beta^r}} = \left(\lambda (\rscri^2-r^2) + 3\alpha^2\chi + \frac{3}{2}\gamma_{rr}{\beta^r}^2 + \frac{9}{2}\sqrt{\gamma_{rr}\chi}\alpha\beta^r \right) \Lambda^r .  
\end{equation}
The resulting outgoing propagation speed is also modified: it is smaller (although still positive) at \scrip (see \fref{fr:lightspeeds}), and it would be positive even inside of the horizon if the $\lambda$ term was not present. The choice of the coefficient in \eref{modifiedshift} giving zero ingoing speed is not unique, but it has been used here for its good behaviour in numerical evolutions, both at the origin and at \scrip. 

Hyperboloidal CMC trumpet initial data (\eref{varshypcompc} and \eref{varsderhypcomp} as derived in \sref{cmcsec}, or any initial data satisfying the relations in \ssref{statiorels}) are a stationary solution of the Einstein equations as described in \sref{formulation}. However, if they are evolved together with gauge conditions whose source functions are filled with hyperboloidal CMC Minkowski data \eref{e:hatvals} as described above, the right-hand-sides of the gauge evolution equations will not be zero. Thus some gauge dynamics will take place in which the trumpet slice readjusts and settles into a new stationary solution, see e.g. the plot on the right of figure 2 in \cite{\procmg}. The change in the slices is easier to understand when depicted as a Carter-Penrose diagram, as in \fref{penboth}. While this scenario is satisfactory in the sense that a long-term solution is found, the initial dynamics does not allow to decouple any potential perturbations of the system from the trumpet dynamics. 
Naively, a way to try to obtain the desired outcome -- trumpet initial data that are a stationary solution of the gauge conditions -- is to fill in the gauge source terms in the gauge conditions with \eref{varshypcompc}, the same data as the one given initially. This has been tested (see section 8.2 in \cite{\thesis}), with the result that a slow exponential growth appeared in the evolutions, causing the simulations to a crash in finite time. The conclusion of these tests is that the chosen trumpet initial data is a stationary but not a stable solution for the gauge conditions with trumpet source terms (more on this in \sref{statio}). Still, the growth in these simulations is slow enough to study small scalar field perturbations, as presented in \cite{\procere}. Whether a different choice of trumpet slice or form of the gauge conditions would not cause the growth is an open question \footnote{There is another potential drawback to this approach: a change in the mass of the BH (for instance, due to some energy that is accreted by it during evolution) would in principle not be taken into account by the source functions, and the gauge conditions may try to force the system into an inappropriate geometry. There is the possibility, at least in spherical symmetry, to evaluate numerically the new value of the BH's mass ``on the fly'' during the evolution, use it to calculate the new trumpet geometry and update the source terms accordingly. An example of this recalculation of the BH's mass and the trumpet is shown for some evolution variables in figures 8.30 and 8.31 in \cite{\thesis}}. 
Meanwhile, an attempt to combine stability and stationarity together is described in \sref{statio}, where a solution for the gauge conditions with hyperboloidal Minkowski source functions is determined for a specific setup. 

\section{Constant-mean-curvature initial data}\label{cmcsec}  

\subsection{Main ingredients of hyperboloidal conformal compactification}

At the core of the hyperboloidal approach is the foliation of spacetime along hyperboloidal slices, which can be characterized as the level sets of a specific parameter. This parameter is taken to be the hyperboloidal time coordinate $t$, and it is related to the usual time coordinate $\tilde t$ via the height function $h(\tilde r)$ \cite{Gentle:2000aq,Malec:2003dq} as 
\begin{equation}\label{trafot}
t = \tilde t-h(\tilde r) .
\end{equation}
The height function satisfies $dh/d\tilde r < 1$ everywhere except asymptotically, where $dh/d\tilde r|_\scri=1$ holds, thus characterizing the hyperboloidal slices as spacelike but extending to \scrip. 

In order to reach future null infinity with a finite value of the spatial coordinates, the radial coordinate $\tilde r$ on a hyperboloidal slice is compactified into a new $r$ using a compactification factor $\aconf(r)$ 
\begin{equation}\label{trafor}
\tilde r=\frac{r}{\aconf(r)} .
\end{equation}
Following \eref{rescmetric}, the line element is conformally rescaled by the conformal factor $\Omega$, to provide regular metric components all the way to \scrip
\begin{equation}\label{confresc}
ds^2 = \Omega^2d\tilde s^2 .
\end{equation}
The compactification factor $\aconf$ is not to be confused with the conformal factor $\Omega$, as they are a priori different quantities. While the conformal compactification method relies in both having the same (or at least proportional) behaviour near \scrip, their behaviour in other parts of the domain (especially at the location of the BHs) can be chosen to be very different. For the spherically symmetric data considered here, an example of this is illustrated in \fref{omegas}. 

\subsection{Spherically symmetric conformally compactified hyperboloidal slices}

A suitable starting point to derive spherically symmetric vacuum initial data on a hyperboloidal slice is the following line element on an uncompactified Cauchy slice, 
\begin{subequations}\label{ein:lielphysboth}
\begin{eqnarray}\label{ein:lielphys}
d\tilde s^2 &=& -A(\tilde r)d\tilde t^2+\frac{1}{A(\tilde r)}d\tilde r^2+\tilde r^2 d\sigma^2 \\ &=&  -A(\tilde r)dt^2 -2A(\tilde r)h'(\tilde r)dt d\tilde r+\frac{1-A(\tilde r)\left(h'(\tilde r)\right)^2}{A(\tilde r)}d\tilde r^2+\tilde r^2 d\sigma^2 , \label{ein:lielphysh}
\end{eqnarray}
\end{subequations}
expressed first in terms of the usual time $\tilde t$, and then in terms of the hyperboloidal time coordinate $t$ after using \eref{trafot}. 
This ansatz for the initial metric is general enough to consider flat spacetime, the Schwarzschild and Reissner-Nordstr\"om (RN) spacetimes, and the addition of a non-vanishing cosmological constant.
After applying the radial compactification \eref{trafor} and conformal rescaling \eref{confresc}, it becomes
\begin{equation}\label{fsthyp}
ds^2= -A\,\Omega^2dt^2+\frac{\Omega^2}{\aconf^2}\left[-2A\,h'\,(\aconf-r\,\aconf')dt\,dr+\frac{\left[1-\left(A\,h'\right)^2\right]}{A}\frac{(\aconf-r\,\aconf')^2}{\aconf^2}d r^2 + r^2 d\sigma^2\right] ,
\end{equation}
where $A$ and $h'$ are functions of $\frac{r}{\aconf}$, while $\aconf$ and $\Omega$ depend on $r$.

\subsection{Hyperboloidal CMC slices}\label{ss:htf}

A convenient way of slicing spacetime is doing it along constant-mean-curvature (CMC) slices, on which the trace of the physical extrinsic curvature ($\tilde K$) is a constant. A special case of CMC slices are maximal slices \cite{PhysRevD.7.2814}, where $\tilde K =0$ and spatial infinity is reached asymptotically. Maximal Schwarzschild trumpet slices are analytically described in \cite{Baumgarte:2007ht}. Generalizations for slices with a non-vanishing $\tilde K$ are considered in \cite{Iriondo:1995ar,Gentle:2000aq,Malec:2003dq,Malec:2003dr} and including the critical case for trumpets in \cite{Malec:2009hg,Buchman:2009ew}, while a study of CMC slices in the RN geometry has been performed in \cite{Tuite:2013hza}. 

This derivation of a height function providing CMC slices follows \cite{Gentle:2000aq,Malec:2003dq}. See e.g. subsection 3.2.2 in \cite{\thesis} for a more detailed derivation. The basic procedure is to express the unit normal $\tilde n^a$ to the hypersurface in terms of the metric \eref{ein:lielphysh} and use it to calculate the expression for the trace of the physical extrinsic curvature 
\begin{equation}
\tilde{ K} = -\frac{1}{\sqrt{-\tilde g}}\partial_a\left(\sqrt{-\tilde g}\,\tilde n^a\right) = -\frac{1}{r^2}\partial_r\left[ r^2 \frac{A^{3/2}(\tilde r)\,\hpt}{\sqrt{1-\left(A(\tilde r)\hpt\right)^2}} \right] .
\end{equation}
Setting it equal to a constant value of $\tilde K=\Kc<0$ and introducing $\Cc$ as an integration constant, the first derivative of the height function is isolated to give 
\begin{equation}\label{hfunc}
\hpt = - \frac{\frac{\Kc\,\tilde  r}{3} + \frac{\Cc}{\tilde r^2} }{A(\tilde r)\sqrt{A(\tilde r)+\left(\frac{\Kc\,\tilde  r}{3} + \frac{\Cc}{\tilde r^2}\right) ^2}}.
\end{equation}
The expression for the flat spacetime case is obtained by setting $A(\tilde r)=1$ and $\Cc=0$, and the height function can be integrated to $h(\tilde r) = \sqrt{(3/K_{CMC})^2+\tilde r^2}$.

Comparing our line element of initial data \eref{fsthyp} with our metric ansatz \eref{e:linel} and substituting \eref{hfunc}, we assign the following initial values to our metric components, {where the notation  $X_0\equiv X(t=0)$ is used:} 
\begin{subequations} \label{varshypcomp}
\begin{eqnarray}
&&\gamma_{\theta\theta0} = 1, \quad \chi_0 = \frac{\aconf^2}{\Omega^2} , \quad
\gamma_{rr0} =\frac{(\aconf-r\,\aconf')^2}{\tilde\alpha_0^2 \ \aconf^2} , \quad \alpha_0 = \Omega\ \tilde\alpha_0  , \label{grrini} \\
&& \beta^r_0 = \frac{\left(\frac{\Kc\,r}{3\aconf}+\frac{\Cc\aconf^2}{r^2}\right)\tilde\alpha_0 \, \aconf^2}{(\aconf-r\,\aconf')}, \quad\textrm{with}\quad \tilde\alpha_0=\sqrt{A(\frac{r}{\aconf})+\left(\frac{\Kc\,r}{3\aconf}+\frac{\Cc\aconf^2}{r^2}\right)^2}.  \label{betarini}
\end{eqnarray}
\end{subequations}

A height function determined by imposing CMC is not the only suitable choice. It could also be only asymptotically CMC \cite{Schinkel:2013zm}. Other possible options are e.g. \cite{Hilditch:2016xzh}, where $h'(\tilde r)$ is chosen to provide unit outgoing radial coordinate lightspeed, and \cite{PanossoMacedo:2019npm}, where $h(\tilde r)$ is introduced as part of the minimal gauge \cite{Ansorg:2016ztf}. 

\subsection{Hyperboloidal CMC trumpet slices}

The choice of the integration constant $\Cc$ is a relevant matter, as a critical value exists that provides trumpet \cite{Hannam:2006vv} CMC data \cite{Buchman:2009ew} in an equivalent way as done in the non-hyperboloidal case \cite{Baumgarte:2007ht}.
This critical value of $\Cc$ depends on $M$, $Q$ and $K_{CMC}$ and is calculated \cite{PhysRevD.7.2814,Baumgarte:2007ht} by setting to zero the discriminant of the denominator (6th order polynomial) of \footnote{The initial ansatz can be as general as $A(\tilde r) = 1-\frac{2M}{\tilde r}+\frac{Q^2}{\tilde r^2}+\frac{\Lambda}{3}\tilde r^2$, but in this work only the case with vanishing cosmological constant is considered.}
\begin{equation}\label{poly}
\gamma_{\tilde r\tilde r\,0}(\tilde r) = \frac{1}{A(\tilde r)+\left(\frac{K_{CMC}\,\tilde r}{3}+\frac{C_{CMC}}{\tilde r^2}\right)^2}.
\end{equation}

\begin{figure}[htbp!!]
\center
\includegraphics[width=0.6\linewidth]{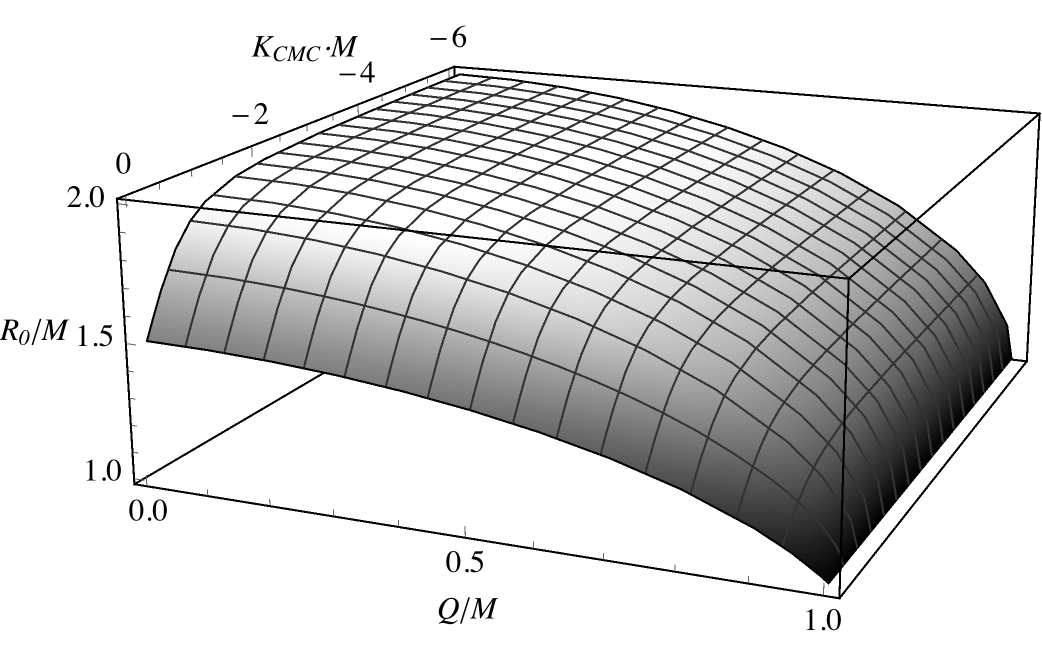}\vspace{-2ex}
\caption{Innermost value of the Schwarzschild-like radial coordinate (the double root $R_0$) reached by the outer CMC trumpet slice, as a function of $\Kc$ and $Q$ (for zero cosmological constant). Taken from \cite{\thesis}.}
\label{R0vsKQ}
\end{figure}

With this critical choice of $\Cc$ (see (3.42) in \cite{\thesis} for the explicit expression for RN), the denominator now has a double real root at $\tilde r=R_0$. This finite value of the radial coordinate $\tilde r$ is where the slice (reaching \scrip{} in its outer end) finishes, corresponding to the location of the trumpet \cite{Buchman:2009ew}. However, in terms of proper distance the inner end of the slice is infinitely far away from the singularity.
For instance, in the Schwarzschild case and for maximal $K_{CMC}=0$ the double root is $R_0=3M/2$ \cite{PhysRevD.7.2814,Baumgarte:2007ht}, whereas for $K_{CMC}\to-\infty$ it tends to $R_0\to2M$. The dependence of the double root $R_0$ on the charge $Q$ and the value of $\Kc$ is shown in \fref{R0vsKQ}. Note that in the extreme Reissner-Nordstr\"om case ($Q=M$), $R_0/M$ is always unity. 

\begin{figure}[htbp]
  \begin{subfigure}[t]{0.56\textwidth}
        \includegraphics[width=\linewidth]{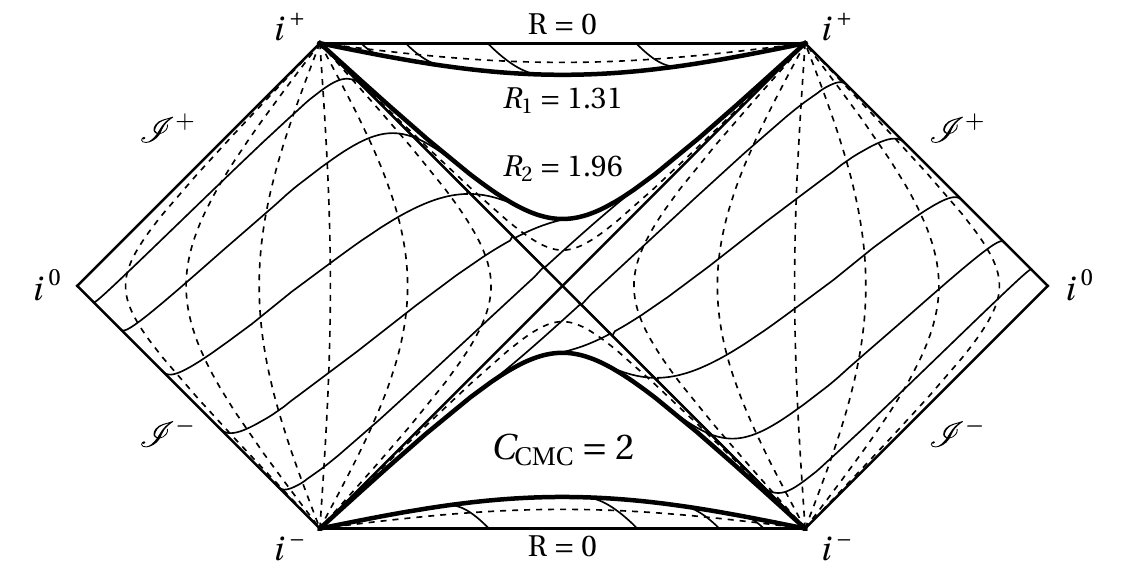}
        \caption{Incomplete slices entering the white hole.}
        \label{fig:pen1}
    \end{subfigure}  
    \begin{subfigure}[t]{0.41\textwidth}
        \includegraphics[width=\linewidth]{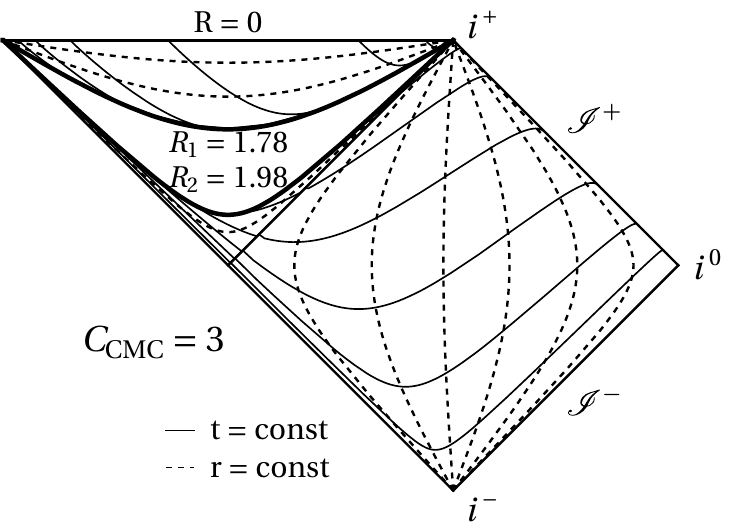}
        \caption{Incomplete slices entering BH.}
        \label{fig:pen2}
    \end{subfigure}  
    \begin{subfigure}[t]{0.43\textwidth}
        \includegraphics[width=\linewidth]{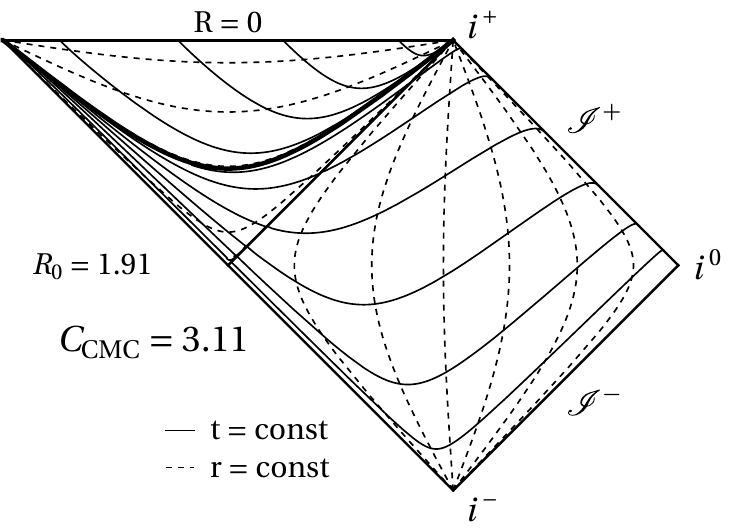}
        \caption{Trumpet slices.}
        \label{fig:pen3}
    \end{subfigure}  
    \begin{subfigure}[t]{.43\textwidth}
        \includegraphics[width=\linewidth]{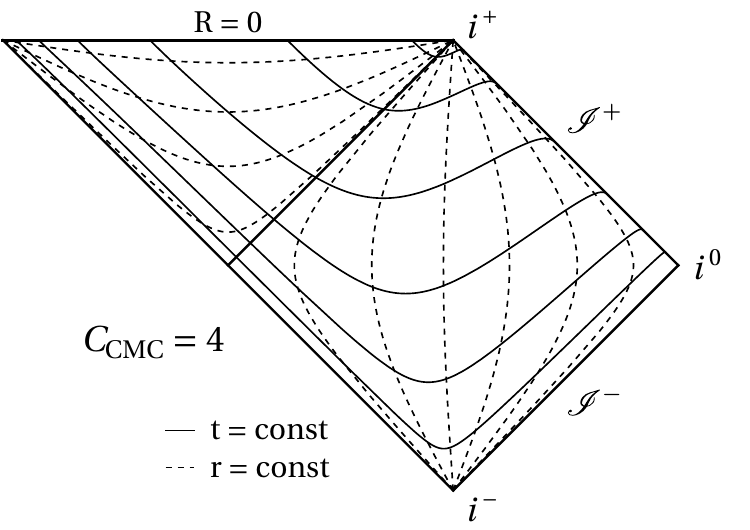}
        \caption{Slices reaching the singularity.}
        \label{fig:pen4}
    \end{subfigure}
    \caption{Carter-Penrose diagrams representing hyperboloidal CMC Schwarzschild foliations with $M=1$ and $K_{CMC}=-1$ and the the choices of $\Cc$ of the height functions depicted in \fref{ploth}. They correct the versions included in figures 3.3 and 3.4 in \cite{\thesis}. In subfigures \ref{fig:pen1} and \ref{fig:pen2}, the height function is imaginary between $R_1$ and $R_2$. For the critical $\Cc$ (\sfref{fig:pen3}), the slices that reach the singularity and those that reach $\scri^+$ are separated by a thick line located at $\tilde r=R_0$ that corresponds to the double root. In the $\Cc=4$ case on \sfref{fig:pen4}, the hyperboloidal slices go all the way from $\scri^+$ into the singularity. This last diagram is qualitatively the same (with different values of the parameters) as the Penrose diagram in figure 10 in \cite{Zenginoglu:2007jw}. The present numerical experiments use the outer slices (those reaching \scrip) of \sfref{fig:pen3}, as the critical value of $\Cc$ allows to map the trumpet slice to the whole of the radial coordinate range $r\in(0,\rscri]$.} 
    \label{hypfol1}
\end{figure}

The effect of $\Cc$'s value on the CMC slices is depicted in figure~1 in \cite{Buchman:2009ew} and in \cite{Tuite:2013hza}, and illustrated in \fref{hypfol1} in the form of Carter-Penrose diagrams of the Schwarzschild spacetime. 
For a value of $\Cc$ smaller than the critical one, the denominator of \eref{poly} has two different real roots $(R_1,R_2)$ for $\tilde r$, the outer one $R_2$ corresponding to the location of the minimal surfaces mentioned in \cite{Buchman:2009ew}. If $\Cc<-\frac{1}{3}\Kc \left(M + \sqrt{M^2-Q^2}\right)^3$ (such as the example shown in \fref{fig:pen1}), the slices reach inside of the white hole, while for a larger value of $\Cc$ (\fref{fig:pen2}) they enter the BH. Quantities become complex for $\tilde r\in(R_1,R_2)$; the corresponding part of the diagrams is left in white. 
As mentioned above, for the critical value of $\Cc$ a double root appears and complete CMC trumpet slices (\fref{fig:pen3}) exist, joining either \scrip to the symmetric point to future timelike infinity ($i^+$) (the outer slices) or the singularity to $i^+$ (the inner ones). 
For $\Cc$ larger than the critical one (\fref{fig:pen4}), there is no root to the polynomial in \eref{poly} and the CMC slices extend between null infinity and the singularity. 
Examples of the outer CMC Schwarzschild trumpet slices for critical $\Cc$ for different values of $\Kc$ is given in figure 3.5 in \cite{\thesis}, showing the maximal case $\tilde K=0$ corresponding to the usual trumpet slices \cite{Baumgarte:2007ht} in the first subfigure. For a positive value of $\Kc$ (in the current sign convention), the hyperboloidal slices reach past null infinity $\scri^-$ instead of \scrip. 
Equivalent Penrose diagrams depicting CMC slices for the RN spacetime (with $A(\tilde r) = 1-\frac{2M}{\tilde r}+\frac{Q^2}{\tilde r^2}$) are shown in figures 3.11, 3.12, 3.13 and 3.14 in \cite{\thesis}. The non-extremal case is illustrated by the choice $Q=0.9M$. The trumpet case (with critical $\Cc$) should compare to panel 2 in figure 2 in \cite{Tuite:2013hza}; the slices have a different profile because theirs were probably not numerically determined. 
The extremal case $Q=M$ is shown in 3.14 in \cite{\thesis} and it has the feature that all slices are trumpet ones with $R_0=M$ always, as can also be seen in \fref{R0vsKQ}. 
For the over-extreme case ($Q>M$) no critical value of $\Cc$ is found, thus no trumpet slices can be constructed (at least with this method). 

\begin{figure}[h!!]
\center
\includegraphics[width=0.75\linewidth]{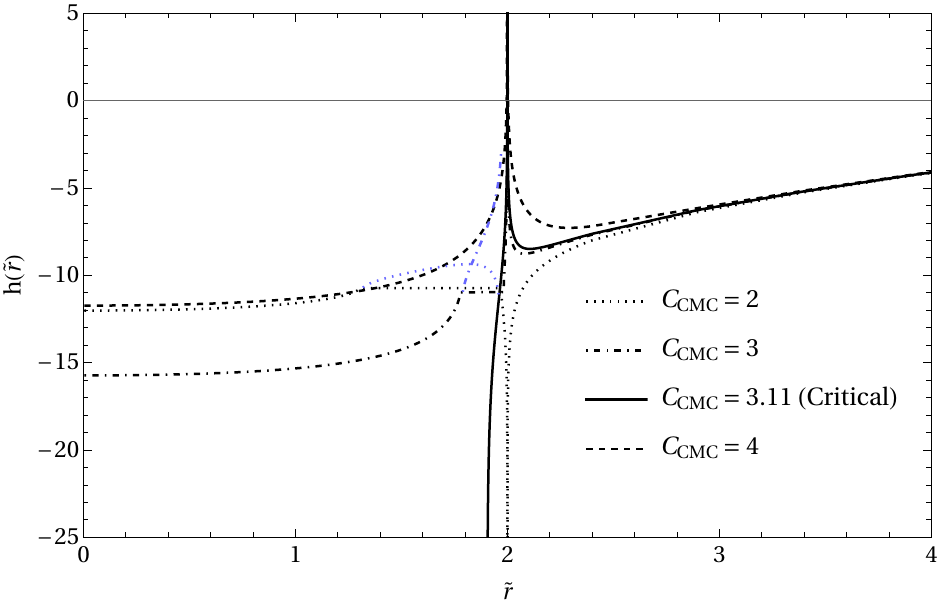}\vspace{-3ex}
\caption{Example of numerically integrated height functions for the Schwarzschild spacetime with $M=1$ and $\Kc=-1$. The values of $\Cc$ correspond to those used in \fref{hypfol1}. The integration constant has been set in such a way that as $\tilde r\to\infty$ the Schwarzschild height function approaches the flat spacetime one ($h(\tilde r) = \sqrt{(3/K_{CMC})^2+\tilde r^2}$). In blue the parts of the height function that are imaginary (for $\Cc$ smaller than the critical one).} 
\label{ploth}
\end{figure}

The effect of $\Cc$ can also be seen in the Schwarzschild examples of the height function $h(\tilde r)$ shown in \fref{ploth}. They were used to construct the respective Penrose diagrams in \fref{hypfol1}. The height function, integrated numerically from \eref{hfunc}, if expressed in Schwarzschild coordinates has a coordinate singularity at the location of the horizon ($\tilde r=2M$): it diverges downward for $\Cc=2$ because it enters the white hole, and upwards for the cases crossing the BH horizon. The region between ($R_1,R_2$) is complex for the subcritical values of $\Cc$ (absence of curves), so it is not straightforward how to join the inner and outer real parts of the corresponding height functions. For the critical case, the outer part of the height function goes to $-\infty$ at the root $\tilde r = R_0=1.905$. 
For the supercritical $\Cc=4$, the height function attains a finite value at the singularity $\tilde r=0$. The integration constants for each case have been set in such a way that as $\tilde r\to\infty$ the Schwarzschild height functions approach the flat spacetime one $h(\tilde r) = \sqrt{(3/K_{CMC})^2+\tilde r^2}$.

Initial data developed in \cite{Buchman:2009ew,Schinkel:2013tka,Schinkel:2013zm} aims for its evolution using excision. For that purpose, any CMC slices with $\Cc>-\frac{1}{3}\Kc \left(M + \sqrt{M^2-Q^2}\right)^3$ (intersecting the BH horizon), are probably suitable, as in any case they will be cut before reaching the singularity. 
The puncture approach pursued here however compactifies the full slice, so that the most suitable choice is the CMC trumpet slice for the critical value of $\Cc$. This is the case that will be considered from now on. 

\subsection{Compactification $\aconf$ and conformal rescaling $\Omega$ for CMC slices} \label{ss:compact}

\begin{figure}[htbp!!]
\center
	\includegraphics[width=0.53\linewidth]{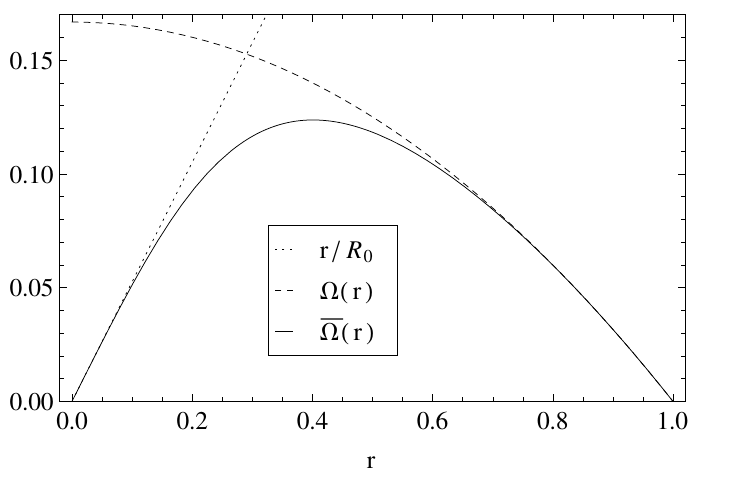}
\caption{Conformal ($\Omega$) and compactification ($\aconf$) factors, as well as the behaviour of $\aconf\sim r/R_0$ near the origin (with $R_0$ the location of the trumpet in uncompactified Schwarzschild radial coordinate), for $K_{CMC}=-1$ and critical $\Cc$. This figure is also included in \cite{\procere}.}\label{omegas}
\end{figure}

It is convenient to determine the compactification factor $\aconf$ by imposing a conformally flat initial spatial metric, which is in a way equivalent to transforming to the isotropic radial coordinate, 
\begin{equation}\label{e:confflat}
\gamma_{rr\,0} = \frac{(\aconf-r\aconf')^2}{\aconf^2\left[A(\frac{r}{\aconf})+\left(\frac{K_{CMC}\,r}{3\aconf}+\frac{C_{CMC}\aconf^2}{r^2}\right)^2\right]} \equiv 1 ,
\end{equation}
as this is also a simple choice compatible with an initial zero \eref{Lambdarstatio}. The factor $\aconf$ is expected to vanish at the same rate as the conformal factor $\Omega$ at $\scri^+$, but it is allowed to have a different behaviour elsewhere.
Expression \eref{e:confflat} is solved numerically for $\aconf$; a suitable procedure is described in subsection~6.6.1 of \cite{\thesis}. 
The result for the Schwarzschild case is shown as the solid line in \fref{omegas}, with the dotted line representing the linear behaviour of $\aconf$ near $r=0$, corresponding to the asymptotic behaviour near the trumpet $\tilde r=R_0$ (see (5) in \cite{Baumgarte:2007ht}). The resulting $\aconf$ will only extend to $r=0$ if the critical value of $\Cc$ is used. If $\Cc$ is larger than the critical value, $\aconf$ will diverge {as $r\to0$}, while using a value smaller than the critical $\Cc$ will provide a $\aconf$ that does not reach the origin (and does not vanish at the smaller $r$ it gets to). 

In flat spacetime ($A(\frac{r}{\aconf})=1$), condition \eref{e:confflat} can be solved analytically resulting in expression \eref{ein:omega}, where $\aconf$ is substituted by $\Omega$. This result is commonly seen in the literature \cite{Husa:2002zc,Schneemann}, and has been used in preceding work \cite{\pap,\papgauge} as conformal and compactification factors for the flat spacetime case.
In order to ensure that the slices reach \scrip (instead of ending at spacelike infinity $i^0$), \eref{ein:omega} satisfies that $\Omega|_\scri=0$ and $\nabla_a\Omega|_\scri\neq0$. Then, its behaviour at $\scri^+$ is unaffected by the choices of the parameters $M$ and $C_{CMC}$ and it is well-behaved at the origin of the coordinate system. It is thus a good candidate to be used as a time-independent conformal factor $\Omega$ in general, so this is indeed also the choice in this work dealing with BH spacetimes, as already mentioned towards the end of \ssref{sphersym}. 
The profile of this choice of $\Omega$ appears in \fref{omegas} as a dashed line. 

\begin{figure}
    \centering
    \begin{subfigure}[t]{0.48\textwidth}
        \includegraphics[width=\linewidth]{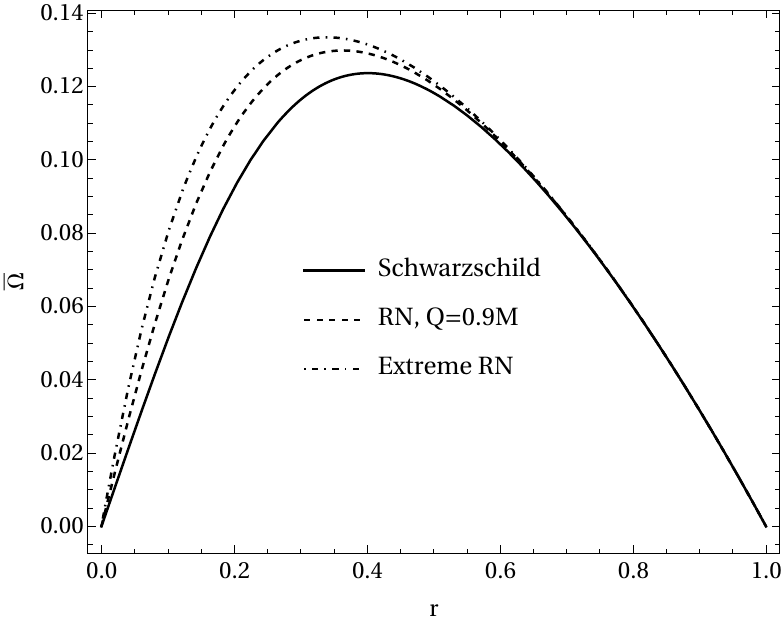}
        \caption{CMC trumpet compactification factor.}
        \label{aconfs}
    \end{subfigure}  
    \begin{subfigure}[t]{0.48\textwidth}
        \includegraphics[width=\linewidth]{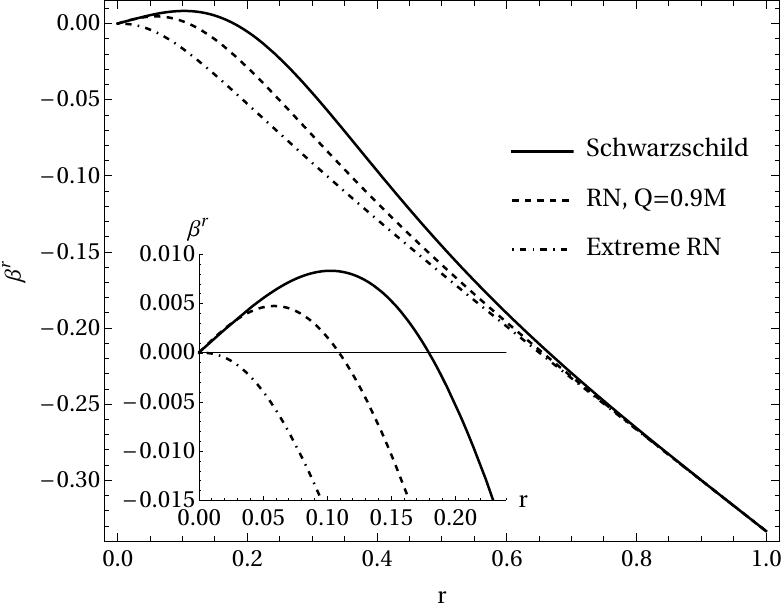}
        \caption{CMC trumpet values for shift $\beta^r$.}
        \label{betarcmc}
    \end{subfigure}
\caption{A version of these figures was originally included in \cite{\thesis}. \textbf{(A)} Profiles of the compactification factor $\aconf$ with $K_{CMC}=-1$ and critical $\Cc$ for Schwarzschild, RN with $Q=0.9M$ and extreme RN CMC trumpet geometries. \textbf{(B)} Trumpet values for the shift $\beta^r$ for Schwarzschild, RN with $Q=0.9M$ and extreme RN geometries. The detail in $\beta^r$'s plot shows how the shift in the extreme RN case is never positive, a consequence of the trumpet and the horizon being mapped to the same point of the compactified radial coordinate.}
\label{schwrn}
\end{figure}

A comparison between compactified spherically symmetric BHs for various values of the charge $Q$ is shown in \fref{schwrn}. In \fref{aconfs}, the CMC trumpet compactification factors $\aconf$ corresponding to Schwarzschild, to RN with $Q=0.9M$ and to extreme RN ($Q=M$) are presented. 
An interesting effect of the extremality of the $Q=M$ case is that the cylindrical infinity of the trumpet {\it and} the BH horizon are mapped to the same point $r=0$ of the isotropic radius $r$. This can be easily recognised by looking at the profiles of the CMC trumpet initial values of the shift $\beta^r$, displayed on the right in \fref{betarcmc}. In the Schwarzschild and non-extreme RN cases the shift is positive at the horizon (mapped to $r_{Schw}\approx0.13$ and $r_{RN+}\approx0.071$ for $Q=0.9M$ respectively), but in the extreme case the shift never becomes positive.
The curves corresponding to the Schwarzschild case of the shift $\beta^r$ in \fref{betarcmc} (see plot on the left figure~3.10 in \cite{\thesis} for a representation of the lapse) compare to figure 5 in \cite{Hannam:2006xw} and figure 2 in \cite{Baumgarte:2007ht}, with the difference that here the data are compactified on a hyperboloidal slice instead of a spacelike one (with vanishing $\tilde K$).

\subsection{Vacuum initial data: Schwarzschild spacetime}

\begin{figure}[htbp!!]
\center
	\includegraphics[width=0.99\linewidth]{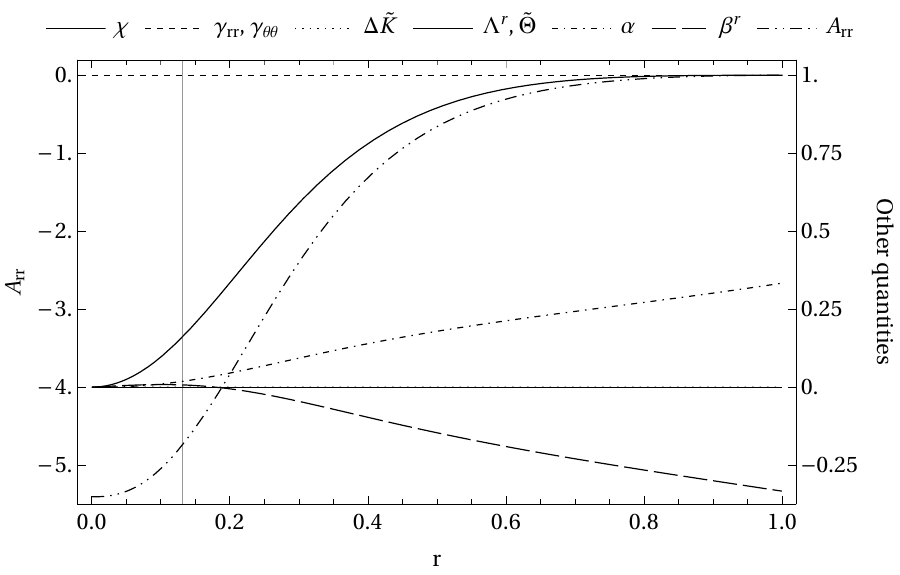}
\caption{CMC Schwarzschild trumpet BH initial data for the evolution variables with $\Kc=-1$, $M=1$ and critical $\Cc$. The vertical line denotes the location of the horizon. As is expected for trumpet data, the lapse $\alpha$ and spatial conformal factor $\chi$ are zero at the puncture. The shift component $\beta^r$ is positive at the horizon and negative at \scrip. Compare to solved-for stationary data shown in \fref{fig:1}.} \label{initialdata}
\end{figure}

From now on, the described CMC trumpet BH initial data, suitable for evolutions using the formalism in \sref{formulation}, will be restricted to the Schwarzschild ($A(\tilde r) = 1-\frac{2M}{\tilde r}$) spacetime. The Reissner-Nordstr\"om case works equivalently. The consideration of a non-vanishing cosmological constant is left for future work. 
After imposing conformal flatness \eref{e:confflat}, $\aconf'$ can be isolated from there and introduced into \eref{varshypcomp}, yielding for the metric components
\begin{subequations}\label{varshypcompc}
\begin{eqnarray}
&&\chi_0 = \frac{\aconf^2}{\Omega^2} , \quad \gamma_{rr0} = \gamma_{\theta\theta0} = 1, 
\\&& \alpha_0 =   \Omega\sqrt{\left(1-\frac{2M\aconf}{r}\right)+\left(\frac{K_{CMC}\,r}{3\aconf}+\frac{C_{CMC}\aconf^2}{r^2}\right)^2} , \quad
\beta^r_0 =  \frac{K_{CMC}r}{3}+\frac{C_{CMC}\aconf^3}{r^2}  ,
\end{eqnarray}
\end{subequations}
and for the extrinsic curvature ($\tilde K=\Kc$) and rest of the variables (calculated from \eref{dKstatio}, \eref{Arrstatio} and \eref{Lambdarstatio} using\eref{varshypcompc})
\begin{equation}\label{varsderhypcomp}
\Lambda^r_0=\tilde\Theta_0=Z_{r0}=0, \quad A_{rr0}=-\frac{2C_{CMC}\aconf^3}{r^3\Omega} \quad\textrm{and}\quad \K_0=0 .
\end{equation}
The background metric $\hat\gamma_{ab}$ is chosen to be the initial value of the evolved one, that is, conformally flat $\hat\gamma_{rr}=\hat\gamma_{\theta\theta}=1$ as indicated in \eref{e:hatvals}.
The profiles of the initial values of the variables \eref{varshypcompc} and \eref{varsderhypcomp} are shown in \fref{initialdata}.
As mentioned in \ssref{ss:gaugebh}, CMC initial data \eref{varshypcompc} and \eref{varsderhypcomp} are a stationary solution of the Einstein equations, but not of the gauge conditions described in \sref{gauge}. 

\section{Stable stationary initial data for given gauge conditions}\label{statio}

Here ``stationary'' initial data means data that correspond to a stationary solution of the Einstein equations and also of the chosen gauge conditions. Only if all RHSs (right-hand sides) of the evolution equations are zero for the initial data, the solution will be stationary. This is valid for both physical and gauge dynamics. 
In \sref{cmcsec} an example was given describing how CMC initial data were a stationary solution of the gauge conditions with CMC-constructed source terms. However, the evolution would start diverging from those data as soon as the simulation started, and variable values would grow exponentially, leading the simulation to crash. This stationary solution was unstable under the small discretization errors naturally arising in a numerical code. 

In this same context, ``stable'' describes an ``attractor-type'' solution to the system. As explained in \sref{cmcsec}, gauge source functions calculated from CMC Minkowski data give a stable stationary end state (after some trumpet gauge dynamics) for some choices of gauge conditions and parameters. Initial and final trumpet states of an instance of that evolution are included in figure~2 in \cite{\procmg}), where the latter also coincides with the final state of a collapsed scalar field creating a BH with the same total mass\footnote{The profiles of the evolution variables at several instances of time during a collapse simulation are shown in figure 8.14 in \cite{\thesis}.} (for the same gauge configuration used). 
Ideally we want stable stationary hyperboloidal trumpet initial data, which will remain a solution of the system even when small initial perturbations are present. 

The way to set up initial data for a given spherically symmetric Schwarzschild BH spacetime (for some chosen values of $M$ and $\Kc$) is to impose the conditions (listed in \ssref{statiorels}) that satisfy the Einstein equations in the non-dynamical regime, and then solve the gauge conditions for stationarity. The latter means setting $\dot\alpha=0$ and $\dot\beta^r=0$, which correspond respectively to the choice of hyperboloidal trumpet slicing and the compactification factor, and solving for the remaining degrees of freedom.
There are several options to tackle the last part: 
\begin{itemize}
\item Solve first the slicing condition on the uncompactified domain (using the uncompactified radial coordinate $\tilde r$ or equivalent) and afterwards use the shift condition to determine a suitable compactification for the radial coordinate, in the form of \eref{trafor}. 
The advantages of this approach are that the steps are performed separately and only one equation is to be solved at a time. The disadvantages are that the first integration is to be performed up to infinite values of the uncompactified radial coordinate, which will introduce considerable errors near \scrip unless some type of compactification is performed, and it also requires determining the location of the trumpet, which is not trivial. This optional partial compactification is difficult to deal with in the second step (imposing stationarity on the shift condition to obtain the compactification), as the solving procedure has to be built on top of it consistently -- all of this assuming that a solution to the equation indeed exists. 
\item Solve both slicing and shift conditions (e.g. in the forms \eref{adotsolve} and \eref{eg:integGammadriver}) for stationarity at the same time. This would in principle allow us great freedom in the gauge conditions that we choose to solve for, provided they give an existing final stable stationary solution for a trumpet after some gauge dynamics. However, there has been no success so far despite numerous attempts. The main difficulty seems to lie within the form of the shift condition. The stable stationary trumpet state reached at late times in experiments sometimes shows a non-smooth profile of the field $\Lambda^r$ at \scrip (see fourth panels in figure 8.31 in \cite{\thesis}). Relation \eref{Lambdarstatio} needs to hold in the stationary regime, and maybe that condition is incompatible with the presence of advection terms in the shift condition \eref{eg:integGammadriver}. In general, the final stable stationary state is very sensitive to the choice of gauge conditions.
\item Solve the slicing condition for the trumpet geometry and impose an initial conformally flat metric to obtain the compactification, in an equivalent way as done for CMC data with \eref{e:confflat}. The advantage is that both equations can be solved at the same time (the compactification allowing to solve all the way to \scrip). The slicing condition (here \eref{adotsolve} is considered) is to be chosen carefully, as it plays an important role in the trumpet geometry \cite{Baumgarte:2022auj}. The disadvantage of this approach is that the shift condition needs to be modified in order to keep the obtained initial data stationary, namely either dropping the advection and ``$\eta$'' terms or having its gauge source functions filled in by the stationary solution found. 
\end{itemize}
The three options have been attempted, with only some success for some specific cases in the third way of proceeding. This best choice will be described in \ssref{statiosolve}, and solved for two example configurations. 

\subsection{Comparison of metric quantities in physical and conformal domains}

For the purpose of clarifying the relations between physical and conformally compactified quantities {in the metric}, let us introduce the following ansatz for the spherically symmetric line element in the uncompactified physical domain in terms of the hyperboloidal time $t$ (again with $d\sigma^2\equiv d\theta^2 + \sin^2\theta d\phi^2$)  
\begin{equation}\label{physansatz}
d\tilde s^2 = - \left(\tilde \alpha^2-\tilde X_{\tilde r\tilde r}\,{\beta^{\tilde r}}^2\right) dt^2 + 2\,\tilde X_{\tilde r\tilde r}\,\beta^{\tilde r} dt\,d\tilde r +  \tilde X_{\tilde r\tilde r}\, d\tilde r^2 +  \tilde X_{\theta\theta}\, \tilde r^2\, d\sigma^2 .
\end{equation} 
The values of the metric components for a hyperboloidal slice can be read off by comparing this metric ansatz to \eref{ein:lielphysh} (or for a Cauchy slice if using $\tilde t$ instead and relating to \eref{ein:lielphys}).

For the conformally compactified version, {as will be used in initial data calculations in \ssref{statiosolve} and } relates to the physical one as in \eref{confresc}, set 
\begin{equation}\label{confansatz}
d s^2 = {\Omega^2 d\tilde s^2=} - \left(\alpha^2- X_{rr}\,{\beta^{ r}}^2\right) dt^2 + 2\, X_{rr}\,\beta^{ r} dt\,d r + X_{rr}\, d r^2 + X_{\theta\theta}\,  r^2\, d\sigma^2 .
\end{equation} 
It relates to the line element \eref{e:linel} used in the evolution formalism by $X_{rr}\equiv\gamma_{rr}/\chi$ and $X_{\theta\theta}\equiv\gamma_{\theta\theta}/\chi$, but for convenience and clarity it is written in terms of the $X$ quantities. Its components can be read off from \eref{fsthyp}, as done for \eref{e:linel} in \eref{varshypcomp} after substitution of \eref{hfunc}. 

The relations between the physical and conformally rescaled metric quantities are the following (most listed in (2.39) and (2.69) in \cite{\thesis})
\begin{equation}\label{acresc}
\tilde\alpha = \frac{\alpha}{\Omega}, \quad \tilde \chi = \Omega^2\bar\chi \quad \textrm{with} \  \tilde\chi = \frac{\tilde\gamma_{\theta\theta}}{\tilde X_{\theta\theta}}\quad \textrm{equivalently as in \eref{e:linel} and} \quad \bar \chi = \frac{\chi}{\aconf^2}, \quad \textrm{so that} \quad\tilde\chi=\frac{\Omega^2}{\aconf^2}\chi. 
\end{equation}
{The quantity $\bar\chi$ accounts purely for the conformal rescaling \eref{rescmetric} on the spatial conformal factor. However, $\chi$ is conformally rescaled and also includes the compactification of the radial coordinate \eref{trafor}.}

The shift does not change due to the 4D conformal rescaling, but its radial components change under a transformation in the radial coordinate. The changes for the $X$ metric components include both effects 
\begin{equation}\label{gbcomp}
\beta^{\tilde r} =  \left(\frac{\aconf-r\aconf'}{\aconf^2}\right) \beta^r, \qquad \tilde X_{\tilde r\tilde r} = \frac{\aconf^2}{\Omega^2} \left(\frac{\aconf}{\aconf-r\aconf'}\right)^2 X_{rr}, \qquad \tilde X_{\theta\theta} = \frac{\aconf^2}{\Omega^2}  X_{\theta\theta}.
\end{equation}

\subsection{Relations holding in the stationary regime}\label{statiorels} 

Subsection \ref{statiosolve} will tackle the derivation of stationary initial data in relation to the gauge conditions, from the same starting point as \cite{Ohme:2009gn}. However, here a slicing condition that has been tested experimentally is considered, and the calculations will take place in the conformally compactified domain instead of the physical one. The procedure will require knowing the conditions that stationarity puts on the metric, which is taken to be the Schwarzschild one from now onward. 

Imposing stationarity on the evolution equation of the metric components allows to find the desired time-independent expressions for the trace of the extrinsic curvature, given in \eref{dKstatio} and \eref{Arrstatio}. 
Setting the RHS of (2.82a) in \cite{\thesis} to zero together with the evolved Z4 constraint $\tilde \Theta=0$, gives the following expression for the quantity $\Delta\tilde K=\tilde K -\Kc$ (the variation of the physical trace of the extrinsic curvature with respect to the background value $\Kc$) in terms of the conformally compactified metric components 
\begin{equation}\label{dKstatio}
\Delta\tilde K = -\Kc + \frac{\Omega}{\alpha}\left({\beta^r}'-\frac{3 {\beta^r} \chi '}{2  \chi }+\frac{{\beta^r} {\gamma_{\theta\theta}}'}{{\gamma_{\theta\theta}}}+\frac{{\beta^r} {\gamma_{rr}}'}{2 {\gamma_{rr}}}+\frac{2 {\beta^r}}{r}\right) - \frac{3\beta^r\Omega'}{\alpha}.
\end{equation}
In essence, the relation above is equivalent to (7) in \cite{Ohme:2009gn}, only here different variables are used and the relation holds in the conformally compactified domain. 
{Setting now the RHS of (2.82b) in \cite{\thesis} to zero} provides the following {expression} for $A_{rr}$ {to hold in the stationary regime}
\begin{equation}\label{Arrstatio}
A_{rr} = \frac{1}{3\alpha}\left(\beta^r\gamma_{rr}' - \frac{\beta^r\gamma_{rr}\gamma_{\theta\theta}'}{\gamma_{\theta\theta}} + 2\gamma_{rr}{\beta^r}' - \frac{2\beta^r\gamma_{rr}}{r}\right). 
\end{equation}
This expression will not be used in further derivations, but is provided here for completeness. 
The stationary expression for $\Lambda^r$ in terms of the spatial metric components, obtained from the Z4 constraint (2.81c) in \cite{\thesis} is 
\begin{equation}\label{Lambdarstatio}
\Lambda^r
=\frac{2\hat{\gamma_{\theta\theta}}}{\hat{\gamma_{rr}}\gamma_{\theta\theta} r}- \frac{2}{\gamma_{rr} r} + \frac{\gamma_{rr}'}{2 \gamma_{rr}^2} -  \frac{\gamma_{\theta\theta}'}{\gamma_{rr} \gamma_{\theta\theta}} + \frac{\hat\gamma_{\theta\theta}'}{\hat\gamma_{rr} \gamma_{\theta\theta}} -\frac{\hat\gamma_{rr}'}{2 \hat\gamma_{rr} \gamma_{rr}} 
= \frac{2}{r}\left(\sqrt{\gamma_{rr}}- \frac{1}{\gamma_{rr}}\right) + \frac{\gamma_{rr}'}{\gamma_{rr}^2}.  
\end{equation}
{After the second equality above $\hat\gamma_{rr}$ = $\hat\gamma_{\theta\theta} = 1$ have been set and the substitution $\gamma_{\theta\theta} = \gamma_{rr}^{-1/2}$ has been imposed. The latter is required by the introduction of the spatial conformal factor $\chi$ in the formulation, and it is to be applied to \eref{dKstatio} and \eref{Arrstatio} as well.} The second expression for $\Lambda^r$ in the stationary regime \eref{Lambdarstatio}, with its formal divergence as $r\to 0$, is not straightforward to solve. To find finite values of $\Lambda^r$ for a $\gamma_{rr}$ that neither diverges not goes to zero at the origin fine-tuning is required (with the exception of $\gamma_{rr}=1$ that gives $\Lambda^r=0$). This condition \eref{Lambdarstatio} possibly puts stringent limitations on the possible solutions of the shift equation \eref{eg:integGammadriver}. 
A way to simplify the problem is to, instead of solving for the shift condition, impose $\Lambda^r=0$ initially and accordingly choose conformally flat initial data, as was mentioned at the beginning of the section and will be used in \ssref{statiosolve}. 

The Schwarzschild spacetime is a static solution of the Einstein equations, given by \eref{ein:lielphysboth} with $A(\tilde r) = 1-\frac{2M}{\tilde r}$. The following relations between metric components in the uncompactified physical domain hold: 
\begin{subequations}\label{relsphys}
\begin{equation}
\tilde \alpha^2-{\beta^{\tilde r}}^2\tilde X_{\tilde r \tilde r} = \tilde \alpha^2-\tilde \beta^2 = 1-\frac{2M}{\tilde r}, \end{equation}
\begin{equation}
\tilde X_{rr}=\frac{1}{\tilde\alpha^2}, \qquad 
\tilde X_{\theta\theta}=1. \label{relsphysX}\end{equation}
\end{subequations}
The first condition is e.g. (9) in \cite{Ohme:2009gn} and corresponds to using the Killing lapse and shift. The second one is used in (8) also in \cite{Ohme:2009gn}, introducing
\begin{equation}\label{betadef}
 \tilde\beta = \sqrt{\tilde X_{\tilde r\tilde r}} \tilde\beta^{\tilde r}  \equiv \sqrt{\frac{\gamma_{\tilde r\tilde r}}{\tilde\chi}} \beta^{\tilde r} \quad \textrm{that transforms via} \quad \tilde\beta = \frac{\beta}{\Omega} \quad \textrm{to} \quad \beta = \sqrt{X_{rr}} \beta^r  \equiv \sqrt{\frac{\gamma_{rr}}{\chi}} \beta^r. 
\end{equation}
The last expression in \eref{relsphysX} means that the physical areal radius remains constant. 

The equivalent expressions in the conformally compactified domain, obtained from \eref{relsphys} using the transformation relations in \eref{acresc} and \eref{gbcomp}, are 
\begin{subequations}\label{relsconf}
\begin{equation}\alpha^2-{\beta^{r}}^2\frac{\gamma_{rr}}{\chi}\equiv \alpha^2-{\beta^{r}}^2X_{rr} = \alpha^2-\beta^2 = \const\,\Omega^2\left(1-\frac{2M\bar\Omega}{r}\right),\label{relsconfa} \end{equation}
\begin{equation}\frac{\gamma_{rr}}{\chi}\equiv X_{rr}=\frac{\const}{\alpha^2}\left(\frac{\Omega^2}{\bar\Omega^2}(\bar\Omega-r\bar\Omega')\right)^2, \qquad \frac{\gamma_{\theta\theta}}{\chi}\equiv X_{\theta\theta}=\frac{\Omega^2}{\aconf^2}. \label{relsconfX} \end{equation}
\end{subequations}
The compactification factor $\aconf$ {cannot be chosen freely} (the conformal factor $\Omega$ {can}), but it is to be substituted from the second relation in \eref{relsconfX}, consequence of the {physical areal radius chosen to be constant in \eref{relsphysX}}. 
Note the presence of the $\cp$ factor {(completely unrelated to the speed of light)} in the first two relations: it corresponds to a constant rescaling of the hyperboloidal time coordinate, $t\to\cp\,t$, in the same way as in the transformation to hyperboloidal time (55) in \cite{PanossoMacedo:2019npm}. The constant $\cp$ will take a different value in the stationary regime depending on the gauge equations chosen and the values of their parameters. Relations \eref{relsconf} have been checked experimentally in spherically symmetric hyperboloidal trumpet evolutions. The quantity $\cp$ can be absorbed into the value of $\Kc$ (in the conformal factor) in the above expressions, but this approach will not be followed here. It is indeed possible that the $\cp$ rescaling of the time coordinate in the evolution is a consequence of a rescaling of the conformal factor $\Omega$ (chosen to be time-independent) that takes place as a result of the change in the hyperboloidal slices. Understanding this effect is left for future work. 

\subsection{Solving the slicing condition and imposing an explicitly conformally flat metric}\label{statiosolve}

A delicate evolution variable in the BSSN/Z4 formulations is $\Lambda^i$, which is usually set initially to zero corresponding to a choice of conformally flat initial metric, i.e.~$\gamma_{ij}=\eta_{ij}$. As pointed out after introducing \eref{Lambdarstatio} for this spherically symmetric setup where $\gamma_{\theta\theta}=\gamma_{rr}^{-1/2}$ holds and $\hat\gamma_{rr}=\hat\gamma_{\theta\theta}=1$, the problem is considerably simplified if $\gamma_{rr}=1$ is set as initial condition, ensuring a well-behaved initial $\Lambda^r=0$. This is indeed a conformally compactified version of isotropic coordinates for puncture data \cite{Bruegmann:2009hof,Baumgarte:2022auj}. Thus, from now on the initial metric will be chosen to be conformally flat explicitly. 
This mimics the procedure done for CMC slices in \sref{cmcsec}, where the compactification factor is also determined imposing conformal flatness via \eref{e:confflat}. The difference is that now the slicing will not be CMC, but determined by stationarity of the slicing condition \eref{adotsolve}. 

\begin{figure}
    \begin{subfigure}[t]{0.45\textwidth}
        \includegraphics[width=\linewidth]{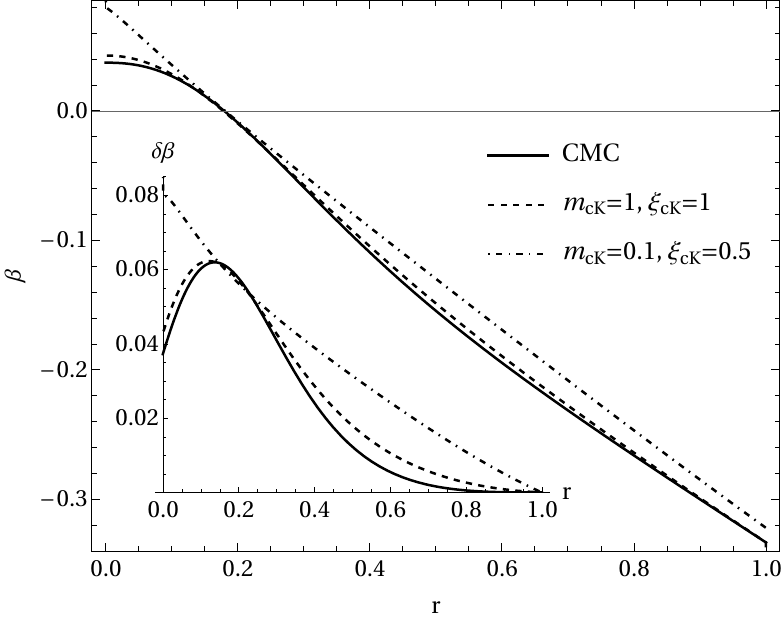}
        \caption{Profiles for $\beta$, with $\delta\beta$ in inset.}
        \label{betasolved}
    \end{subfigure}  
    \begin{subfigure}[t]{0.45\textwidth}
        \includegraphics[width=\linewidth]{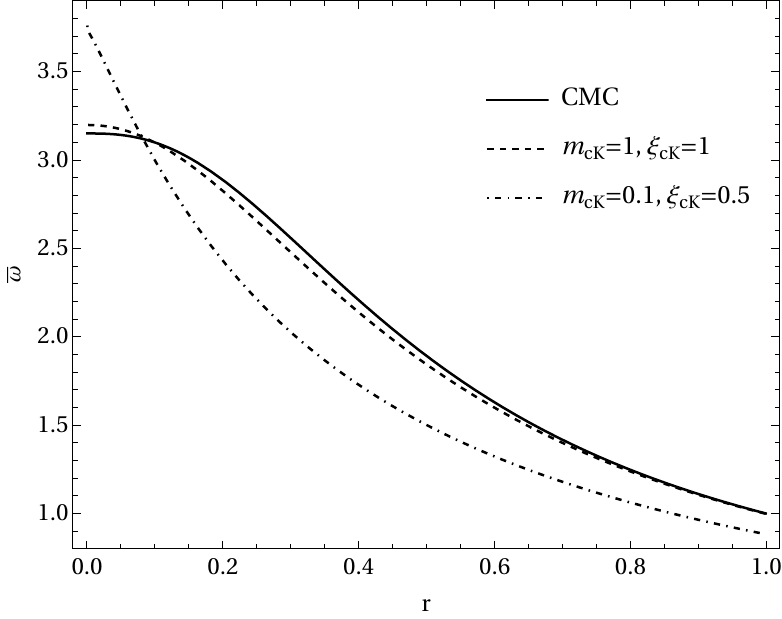}
        \caption{Profiles for $\bar\omega$.}
        \label{wbarsolved}
    \end{subfigure}
\caption{CMC trumpet profiles together with solutions to the slicing condition \eref{adotsolve} and \eref{barwcond} (with minus sign) for two different sets of parameters. The corresponding values of $\cp$ are included in the main text.}
\label{betawbarstatio}
\end{figure}

The coupled system of equations to solve is the left equation in \eref{relsconfX} and \eref{adotsolve}'s RHS set to zero with $\nconst = -m_{cK}\frac{6\rscri}{\Kc}\Omega = m_{cK}(\rscri^2-r^2)$, where $m_{cK}$ is a non-zero constant. The reason why $\nconst$ is set to be proportional to the conformal factor is to ensure that it vanishes at \scrip. In this way the gauge propagation speed associated with the slicing condition will be the physical one at future null infinity (as is the case for the harmonic slicing for $\nconst=0$), as is shown in \fref{fr:lightspeeds}. 
The other relations in \ssref{statiorels} are used to substitute all quantities in those two equations in terms of $\beta$ (defined in \eref{betadef}), the following rescaling of the compactification factor
\begin{equation}
\bar\omega = \frac{\aconf}{r\,\Omega},
\end{equation}
which is expected to be finite and non-zero everywhere in the integration domain, and the constant $\cp$, which will be used as parameter to shoot-and-match on during the solving procedure. The profiles of $\beta$ and $\bar\omega$ for the CMC case, with $\cp=1$, are depicted in \fref{betawbarstatio}. 
The explicit substitutions to be performed are 
\begin{equation}\label{substatio}
\aconf = r\ \Omega\ \bar\omega, \quad \beta^r = \beta \sqrt{\frac{\chi}{\gamma_{rr}}}, \quad \gamma_{\theta\theta}=\gamma_{rr}^{-1/2}, \quad \gamma_{rr} = 1, \quad \chi = r^2\ \bar\omega^2, \quad \alpha = \sqrt{\beta^2+\const\Omega^2(1-2M\bar\omega\Omega)}. 
\end{equation}
The expression from $\chi$ comes from imposing conformal flatness on the right equation in \eref{relsconfX}. The $\alpha$ has been isolated from \eref{relsconfa}. The reason for choosing to substitute $\alpha$ instead of $\beta$ is that the latter, as can be seen in \fref{betasolved}, changes sign over the compactified domain (the shift $\beta^r$ is positive at the horizon and negative at \scrip) and thus cannot be easily substituted. After the substitutions, the left equation in \eref{relsconfX} reads
\begin{equation}\label{barwcond}
\bar\omega' = \frac{\bar\omega}{\Omega}\left(\pm\frac{\sqrt{\beta^2+\const\Omega^2(1-2M\bar\omega\Omega)}}{\cp\,r}-\Omega'\right). 
\end{equation}
The sign providing the RHS that coincides with CMC's $\aconf'$ from \fref{omegas} for substituted CMC data is chosen (the minus one, in this case). 
The resulting equation from \eref{adotsolve} is much longer and has been included in \aref{eqap} as \eref{eqlong}. Both equations are formally diverging at the trumpet and at \scrip. The ellipticity of the coupled system of equations has not been studied. 
For convenience, the system is solved for 
\begin{equation}\delta\beta = \beta - \frac{\cp\, \Kc\, r}{3}\end{equation}
instead of $\beta$, as the former vanishes at \scrip (see inset in \fref{betasolved}). The explicit form of the conformal factor \eref{ein:omega} is also substituted. 

Using Taylor expansions in the radial coordinate around the trumpet is not suitable, as $\alpha$ may not necessarily be proportional to an integer power of the radial coordinate \cite{Bruegmann:2009hof,Baumgarte:2022auj}. The integration will thus start from \scrip{} towards the trumpet, motivating the change of coordinate $x=1-r$. 
Guesses for the initial values of $\delta\beta$ and $\bar\omega$ at \scrip{} ($x=0$) required to start the shoot-and-match integration from there are obtained by Taylor-expanding the equations around \scrip{} up to first order. The values of the variables and their first derivatives at future null infinity, 
\begin{subequations}\label{expansions}
\begin{eqnarray}
&&\delta\beta|_{\scri^+}\approx 0 + x\frac{3 (c-1) \left(c^2 \cnK \Kc^2-3 c \Kc^2 m_{cK}-3 m_{cK} \left(9 \cnK+\Kc^2\right)\right)}{c \Kc \left(2 \left(3 c^2-2\right) \Kc^2+9 (3 c-4) \cnK\right)} \\
&&\bar\omega|_{\scri^+}\approx \frac{\left(3 c^2-1\right) \Kc^2+9 (c-1) \cnK}{2 c^2 \Kc^2}\left[ 1+ x\frac{1}{c^2 \Kc^2 \left(2 \left(3 c^2-2\right) \Kc^2+9 (3 c-4) \cnK\right)}\cdot\left(6 c^4 \Kc^4+\right.\right. \nonumber\\
&&\left.\left.+36 c^3 \cnK \Kc^2-c^2 \Kc^2 \left(45 \cnK+4 \Kc^2+27 m_{cK}\right)-243 c \cnK m_{cK}+27 m_{cK} \left(9 \cnK+\Kc^2\right)\right)\right]
\end{eqnarray}
\end{subequations}
are obtained by imposing regularity of the expansions at \scrip{} for each power of $x$. 
The system of equations is solved using Mathematica's \texttt{NDSolve} function {on the integration domain $x\in[10^{-5},1)$ and a \texttt{WorkingPrecision} of 50. The starting point $x=10^{-5}$ is chosen to avoid the formal divergence of the equations at \scrip, while the integration is carried out up to almost $x=1$ (at least as close as $x=0.9996$). The starting point needs to look visually very near to future null infinity, and be closer to \scrip than any of the gridpoints in the evolutions (see \sref{evolsec}).} Conditions \eref{expansions} are evaluated at $x=10^{-5}$ for the chosen values of $m_{cK}$ and $\cnK$, and a value for $\cp$ is set with high precision to start the integration. The criteria to determine whether the found solution is good enough is for the lapse $\alpha$ to be zero at the trumpet. In practice when using \texttt{NDSolve}, that translates to obtaining profiles of $\delta\beta$ and $\bar\omega$ that look smooth and do not diverge at $x=1$ -- given the difficulty of the integration, a solution with a very small divergence localised beyond $x=0.999$ is taken as valid. 
Unless $\cp$ is very close to the required value, the solutions very quickly become infinite as being integrated towards the origin, given the formally divergent character of the equations. On top of that, there are some regions in the potential values of $\cp$ where the solutions become complex, or they cannot be integrated any closer to the origin than a certain point. That point may correspond to the hyperboloidal equivalent of the ``critical point'' mentioned in \cite{Bruegmann:2009hof,Baumgarte:2022auj}. Besides, the level of fine-tuning required for $\cp$, so that the solution is well-behaved up to $x=1 (\equiv r=0)$, is very high. {What is meant with ``fine-tuning'' for $\cp$ is: the system of equations is solved for a specific value of $\cp$ and, depending on the direction in which the obtained profiles for $\delta\beta$ and $\bar\omega$ are diverging, the next value of $\cp$ is selected to decrease the divergence. This procedure is repeated until a value for $\cp$ that provides regular profiles for the solutions at the origin (and $\alpha$ close enough to zero there) is found.} This makes usual methods to choose the values for $\cp$ in the shooting-and-matching difficult to employ successfully, so that after a careful study of each setup (for chosen values of $m_{cK}$ and $\cnK$), a manual tuning of $\cp$ has been used. 

The profiles of $\beta$ and $\bar\omega$ for two different parameter choices, with the CMC equivalent for comparison, are displayed in \fref{betawbarstatio}. The two configurations considered were (with $M=1$ and $\Kc=-1$): $m_{cK}=1$ and $\cnK=1$, requiring $\cp=0.9996723791389223$, and $m_{cK}=0.1$ and $\cnK=0.5$, with $\cp=0.9661803490$. These values of the parameters were chosen because they {provide a long-term stationary solution} in evolutions of the Einstein equations {(see \ssref{nsevol} for further comments)}. {The number of significant digits required for the value of $\cp$ depends on the method and precision used.}
As can be seen in the plots, the larger $m_{cK}$, the more ``CMC-like'' the slices look near the origin, whereas it has been found experimentally that $\cnK$ has more of an impact in the region near \scrip, allowing solutions to be further from the CMC profile for smaller values of $\cnK$. This is also the behaviour seen in evolutions of CMC trumpet initial data with those parameter choices for the slicing condition. 

\begin{figure}
    \begin{subfigure}[b]{0.315\textwidth}
        \includegraphics[width=\linewidth]{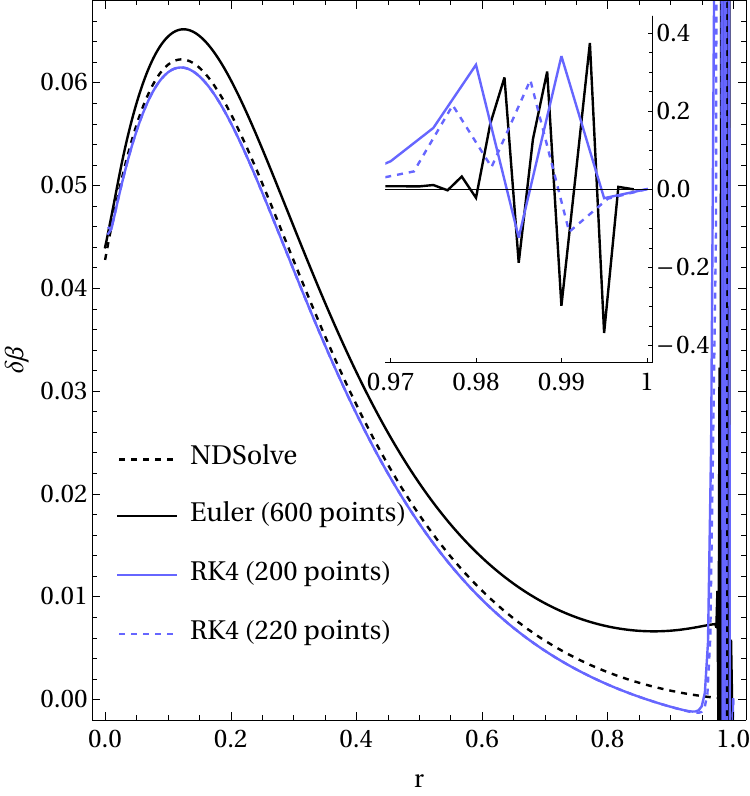}
        \caption{Solutions for $\delta\beta$.}
        \label{dbetafort}
    \end{subfigure}  
    \begin{subfigure}[b]{0.315\textwidth}
        \includegraphics[width=\linewidth]{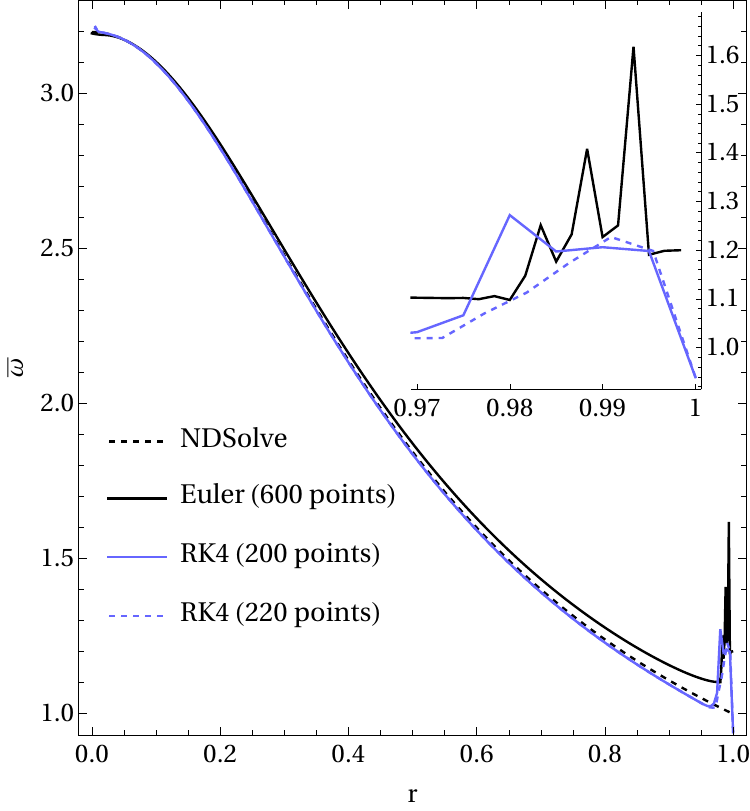}
        \caption{Solutions for $\bar\omega$.}
        \label{wbarfort}
    \end{subfigure}  
    \begin{subfigure}[b]{0.33\textwidth}
        \includegraphics[width=\linewidth]{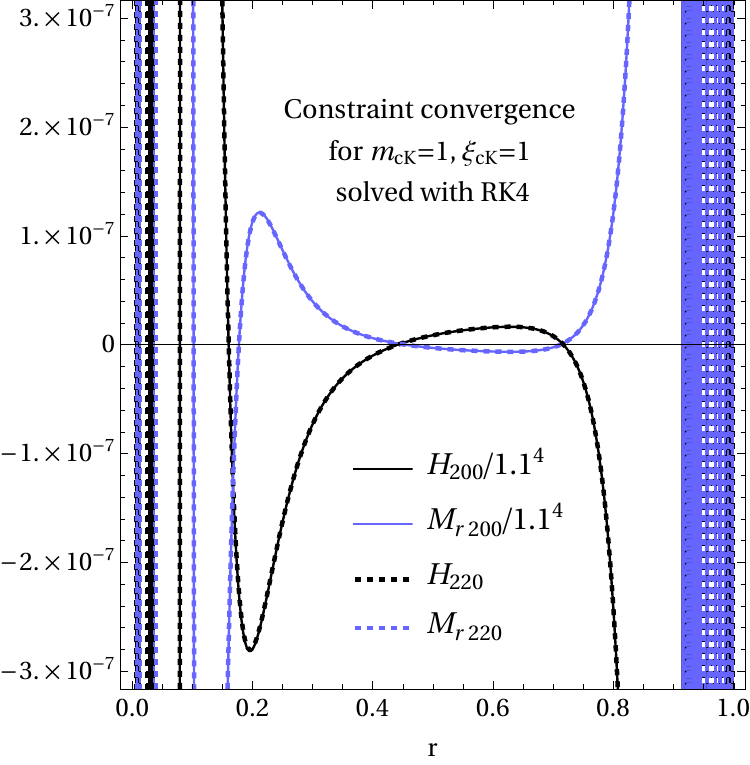}
        \caption{Constraints for RK4.}
        \label{rk4}
    \end{subfigure}
\caption{{{\bf(A)} and {\bf (B)}: Comparison between the \texttt{NDSolve} results (black dashed), and those obtained for a simple Euler solver (solid black) and the Runge-Kutta 4 (blue lines) for the $m_{cK}=1$ and $\cnK=1$ case, specifying the number of gridpoints used for Euler and RK4. The non-smooth part near \scrip for the explicit integration methods (Euler and RK4) is shown in the insets.} Note that the profiles of the solutions show a jump to the left and right of the noisy part. {{\bf(C)} Constraints evaluated for the two solutions obtained with the RK4 integrator (with 200 and 220 points, thus a resolution increase of 1.1). The Hamiltonian and $r$-component of the momentum constraint for the lower resolution are divided by $f^p$, where $f$ is the increase in resolution of 1.1 and $p$ is the order of convergence (4 for RK4). The curves coincide well in the interior of the domain (the solutions converge there), but the noisy part near \scrip shown on subfigures \ref{dbetafort} and \ref{wbarfort} does not converge.}}
\label{fort}
\end{figure}

With \texttt{NDSolve} it is not straightforward to estimate the the error of the solution (even using options like \texttt{AccuracyGoal} or \texttt{PrecisionGoal}), which in turn makes the study of its convergence difficult. {Using $x=\cdot10^{-5}$ as starting point for the integration gave larger residuals, as expected, but this is not enough to systematically study convergence.}
In order to overcome this hurdle, {simple explicit integrators were} implemented to solve the same system of coupled equations. {Those were a 1st order Euler method and a 4th order Runge-Kutta (RK4)}. The explicit integrator was used for the shooting, while a bisection method was used for the matching part (looking how close to zero $\alpha|_r=0$ was for the chosen value of $\cp$). An example of the results obtained for $m_{cK}=1$ and $\cnK=1$ (for a value of $\cp=1.03903643703151$ in the Euler method) is shown in subfigures \ref{dbetafort} and \ref{wbarfort} with a solid black line, also including the \texttt{NDSolve} solution for comparison (black dashed) -- there are obvious differences. 
{Two solutions obtained with the RK4 integrator are also shown in \fref{fort} in blue, solid for 200 points (with $\cp=0.9891442037734391040608175$) and dashed for 220 points ($\cp=0.989160040599287855336$), although they are only distinguishable from each other in the noisy part near \scrip. While these curves are closer to the \texttt{NDSolve} solution, there are still differences between them.} 
The main obstacle in {the explicit integration} methods was that the $\sim 3\%$ of the gridpoints closest to \scrip look very noisy and there is a jump between the values of $\delta\beta$ and $\bar\omega$ at both sides of the noise, even if a solution for the whole domain was found. 
Fine-tuning on the correct value of $\cp$ was difficult {(this is why only two solutions are given for the RK4)}, as the RHS would change sign several times throughout the domain of $\cp$ considered {(quite possibly due to the presence of the noisy part)}, and not all of the potentially promising values would provide a solution extending all the way to the origin. An extra difficulty was that sometimes the RHS would become complex halfway through the integration and no full solution was found, which also happened with \texttt{NDSolve}. 
{Convergence for the part of the RK4 solution between the origin and the noise is shown in \sfref{rk4}: the Hamiltonian and $r$-component of the momentum constraints are evaluated for the 200 and 220-point RK4 solutions and rescaled according to the expected 4th order convergence. Both lines overlap perfectly in most of the interior domain.}

{Apart from the unavoidable fact that the equations are formally singular at the extrema, potential explanations for the delicate tuning of $\cp$ required and the noisy part in the solutions obtained from the explicit integrators are: i) the Taylor expansion \eref{expansions} used to start the integration are not suitable, ii) the slicing condition considered \eref{adotsolve} together with the imposition of conformal flatness do not provide a solution (see \ssref{nsevol} for comments on its stationary solution after evolution).}

\begin{figure}[h!]
\begin{center}
\includegraphics[width=0.99\textwidth]{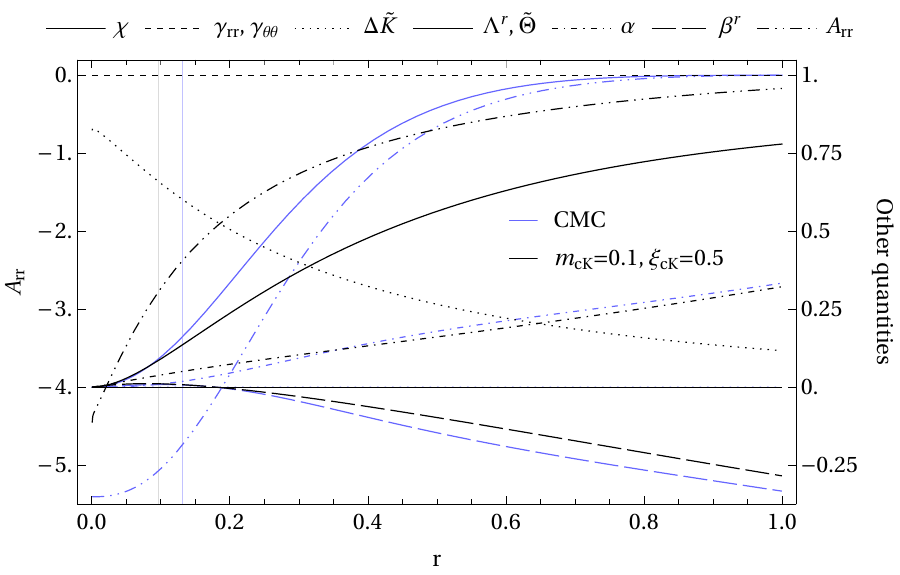}
\end{center}
\caption{Solved-for Schwarzschild trumpet BH initial data for the evolution variables with $\Kc=-1$, $M=1$, $m_{cK}=0.1$ and $\cnK=0.5$ in black, and CMC data shown in \fref{initialdata} in light blue for comparison. The vertical lines denote the respective location of the horizons. The $A_{rr}$ component of the trace-free part of the conformal extrinsic curvature shows a steeper slope at the origin for the solved-for data. The values at \scrip of the corresponding $\chi$, $\alpha$ and $\beta^r$ are different from their CMC values, while the solved-for $\K$ is non-zero in the whole domain. As imposed when constructing both trumpet slices, $\gamma_{rr}=\gamma_{\theta\theta}=1$ and consequently $\Lambda^r=0$.}\label{fig:1}
\end{figure}

Initial data (constructed from the \texttt{NDSolve} solution) for the evolution variables for the $m_{cK}=0.1$, $\cnK=0.5$ case, chosen because its solution differs more from the CMC profile, are presented in \fref{fig:1}. The quantities are calculated from $\beta$, $\bar\omega$ and $\cp$ using relations \eref{substatio}, \eref{dKstatio}, \eref{Arrstatio} and \eref{Lambdarstatio}. The CMC profiles from \fref{initialdata} have also been included in the plot in a light blue color to facilitate comparison. The main qualitative difference between both sets of data is that $\K$ has an positive dependence on $r$ in the non-CMC case. Note that the full trace of the physical extrinsic curvature, $\tilde K = \Kc+\K = -1 +\K$, is still negative everywhere for the solved-for case. The spatial conformal factor $\chi$ is no longer unity at \scrip and $A_{rr}$ reaches the origin (the location of the trumpet) with a steeper slope. 

\begin{figure}
    \begin{subfigure}[t]{0.49\textwidth}
        \includegraphics[width=\linewidth]{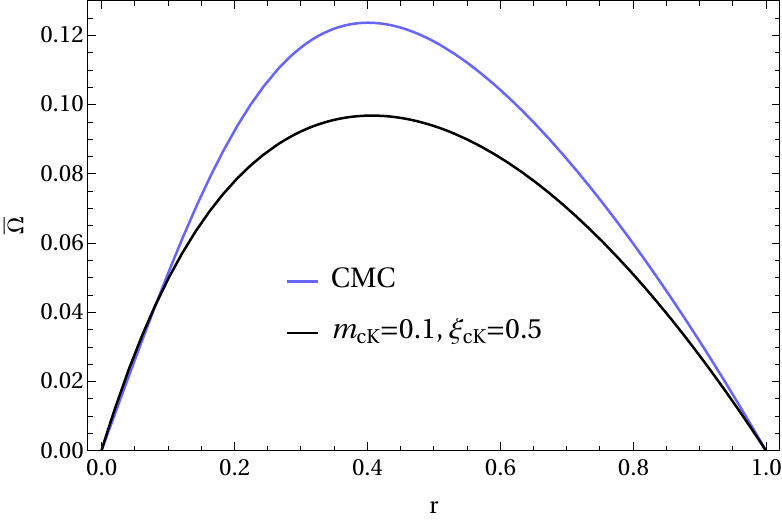}
        \caption{Compactification factor.}
        \label{aconfboth}
    \end{subfigure}  
    \begin{subfigure}[t]{0.49\textwidth}
        \includegraphics[width=\linewidth]{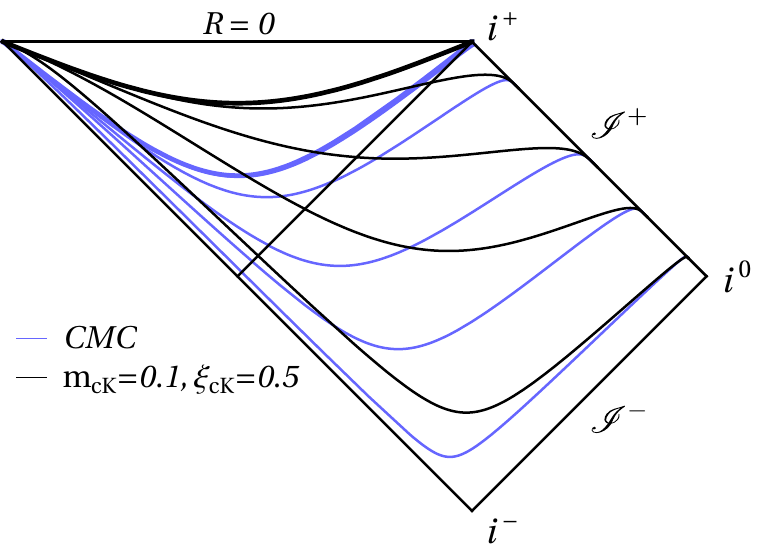}
        \caption{CMC and solved-for slices.}
        \label{penboth}
    \end{subfigure}
\caption{Comparison between CMC trumpet and the solution for $m_{cK}=0.1$ and $\cnK=0.5$. Whereas in the CMC case the trumpet is located at $R_0=1.91M$ of the Schwarzschild radius (as indicated in \fref{fig:pen3}), for the solved-for slices it is at $R_0=1.60M$, closer to the singularity. Accordingly, the slope ($\sim 1/R_0$) near $r=0$ of the compactification factor is steeper. The slices shown on the Penrose diagram correspond to different values of the hyperboloidal time $t$ for CMC and solved-for data; they have been chosen in such a way that the slices coincide at \scrip.}
\label{aconfpenboth}
\end{figure}

The compactification and slicing for the $m_{cK}=0.1$, $\cnK=0.5$ solution, including the CMC profiles for comparison, is shown in \fref{aconfpenboth}. The two compactification factors in \fref{aconfboth} are qualitatively the same. The slope at the origin is different, because the slices of the solved-for solution reach further into the horizon -- the trumpet is located at $R_0=1.60M$. This can be appreciated in \fref{penboth}. 
Full details on the construction of Penrose diagrams for numerical data will be included in \cite{penrosepap}. 

\section{Evolution results}\label{evolsec}

\subsection{Implementation}\label{s:imple}
Simulations are performed with a spherically symmetric code that uses the method of lines with a 4th order Runge-Kutta time integrator and 4th order finite differences, adding Kreiss-Oliger dissipation \cite{kreiss1973methods}. The grid used is staggered (cell-centered), so that it avoids the two points where the equations are formally singular - the origin $r=0$, that corresponds to the value of the Schwarzschild radial coordinate $R_0$ where the trumpet asymptotes to, and $r=\rscri$, which corresponds to future null infinity $\scri^+$. Extrapolating boundary conditions like the outflow boundary conditions in \cite{Calabrese:2005fp} are used at both boundaries.
Off-centered finite difference stencils in the advection terms' derivatives is known to improve stability and performance of numerical relativity simulations \cite{PhysRevD.72.024021,Husa:2007hp,Chirvasa:2008xx}. Thus, advection stencils are up-winded towards larger radii (like on the right in figure 2 in \cite{\papgauge}) where the radial shift component is negative, and down-winded towards smaller radii where the shift is positive. 

The chosen values of the parameters for the simulations considered here are {(unless stated otherwise)}: $\kappa_1=0.5$, $\kappa_2=0$, $\kappa_0=0$, dissipation strength $0.1$, $\Kc=-1$, $m_{cK}=0.1$, $\cnK=0.5$, $\lambda=1$, $\xi_{\beta^r}=0$, $\eta=0,1$. Most simulations use 456 spatial discretization points and a timestep of $dt = 6.667\cdot 10^{-4}$. {The number of spatial gridpoints is enough to resolve the hyperboloidal trumpet initial data considered here, while the timestep is chosen to be below the Courant–Friedrichs–Lewy limit. For some configurations (not in the case of the work presented here), the $dt$ needs to be smaller to account for the presence of stiff terms in the equations.} Evolutions have been performed with the generalized BSSN system.  

As mentioned in the previous section, {convergence of the stationary initial data solutions could so far only be shown on part of the integration domain (see \sfref{rk4}). For those data, noisy features makes the reconstructed initial data for the evolution variables (such as $A_{rr}$ or $\K$) non-smooth enough to pose problems in the evolutions, namely that the simulations crash due to the spiky profiles.} 
Here the focus will be to understand the phenomenological behaviour of the solutions. In any case, any {reasonable} lack of convergence or smoothness of the solved-for initial data will disappear as the evolution progresses, as the gauge conditions will drive the data to the real solution. 
As initial data, the two options depicted in \fref{fig:1} will be used, namely the hyperboloidal Schwarzschild CMC trumpet data \eref{varshypcompc} and \eref{varsderhypcomp}, as well as the solved-for $m_{cK}=0.1$, $\cnK=0.5$ solution {obtained with \texttt{NDSolve}}, which from now on will be called ``\statio'' solution. 
Two different gauge setups will be considered, both using \eref{adotsolve} as slicing condition: the first one will use the integrated Gamma-driver \eref{eg:integGammadriver}, of which neither CMC trumpet data nor the \statio solution are a stationary solution, while the second one will involve the modified shift condition \eref{modifiedshift} without advection terms, whose RHS is zero for both sets of initial data. 

\subsection{Evolution with shift condition including advection terms}

The first test is performed with the Gamma-driver \eref{eg:integGammadriver} shift condition with advection terms and $\eta$ term, together with the slicing \eref{adotsolve}. The expectation is that the trumpet readjustment to happen for the \statio data should be smaller than for the CMC data.

When evolved with $\eta=0$, a small oscillatory behaviour appears in all evolution variables around the initial profiles of the \statio data, no matter if the starting point of the simulation is CMC or \statio data. The amplitude of these oscillations grows slowly (up to the final $t=100$ of these simulations), which indicates that the simulation will crash at some point later in time. 
The damping term with $\eta$ was already introduced in \cite{Alcubierre:2002kk} to avoid strong oscillations in the shift, while other works have found a small value of $\eta$ useful to suppress gauge oscillations, like those affecting eccentricity measurements in \cite{Purrer:2012wy}. Further study of suitable values of $\eta$ and comparison with non-hyperboloidal simulations is left for future work.

\begin{figure}[h!]
\begin{center}
\includegraphics[width=0.99\textwidth]{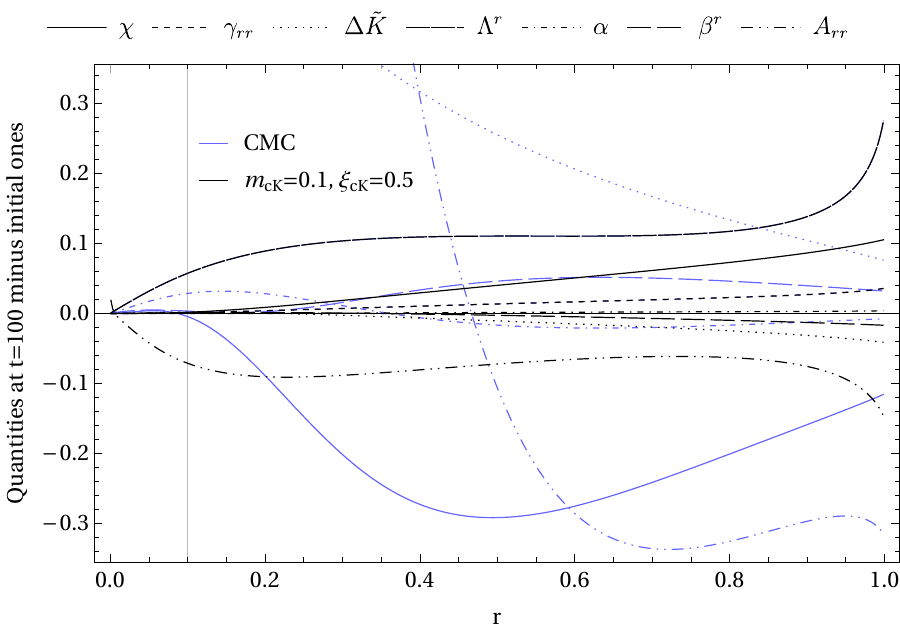}
\end{center}
\caption{In blue are the differences between the initial CMC trumpet data and the final state of the simulation (evaluated at $t=100$), while the black lines show the difference between the $m_{cK}=0.1$, $\cnK=0.5$ initial profile and the final one. The final state is the same for both sets of initial data, as they were both evolved with the same setup: slicing \eref{adotsolve} with $m_{cK}=0.1$ and $\cnK=0.5$, and shift with advection terms \eref{eg:integGammadriver} with $\lambda=1$ and $\eta=1$. The vertical line denotes the final location of the horizon. The differences are larger for CMC initial data. See main text for further details.}\label{sseta}
\end{figure}

If setting $\eta=1$, the dynamics drives the initial data to a stable solution. The differences between the initial and the final profiles are shown in \fref{sseta}. As expected, those differences are larger for CMC initial data, especially the change in $A_{rr}$, which gets up to $2.4$ (beyond the range shown in the figure). The two lines for $\Lambda^r$ lie on top of each other, as both their initial and end states are the same. While the \statio initial data has indeed the advantage that it has required less trumpet dynamics in the evolution, the profiles of the $A_{rr}$ and $\Lambda^r$ quantities near \scrip look slightly diverging, which may cause convergence problems in general {(see comments in next subsection)}. 
{The depicted changes in $\alpha$ show that its final state is smaller than the CMC initial profile, but slightly larger than the \statio one.}
No further study of the parameter space has been performed, because anyway it is not yet clear what gauge conditions are best suited for the hyperboloidal setup. 

\subsection{Evolution with shift condition without advection terms}\label{nsevol}

\begin{figure}
    \begin{subfigure}[t]{0.49\textwidth}
        \includegraphics[width=\linewidth]{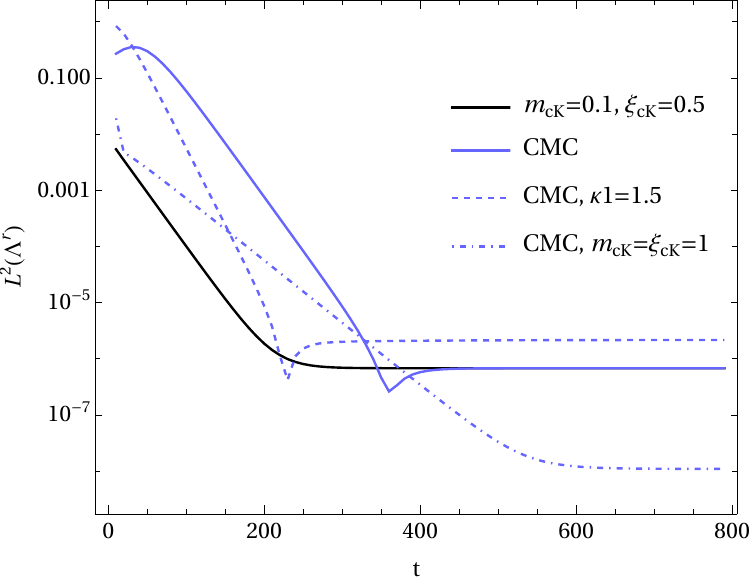}
        \caption{$L^2$-type norm for $\Lambda^r$ over time.}
        \label{nsL2}
    \end{subfigure}  
    \begin{subfigure}[t]{0.49\textwidth}
        \includegraphics[width=\linewidth]{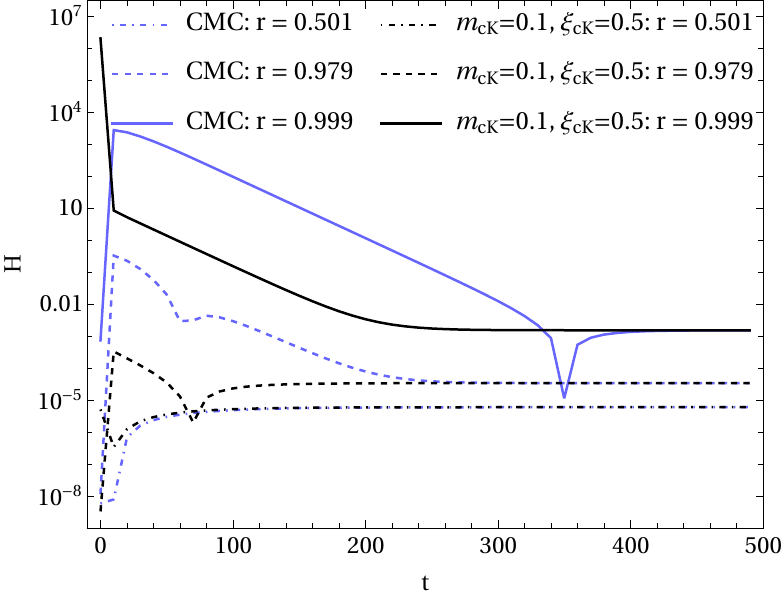}
        \caption{Hamiltonian constraint at several $r$.}
        \label{nsH}
    \end{subfigure}
\caption{Long-term behaviour of unperturbed simulations with advection-term-free shift condition. \textbf{(A)} $L^2$-type norm for $\Lambda^r$'s gridfunctions at given times, $\sqrt{\sum_{i=1}^N(\Lambda^r_i)^2}$ where $i\in[1,N]$ denotes the spatial points. As $\Lambda^r=0$ is the stationary solution of \eref{modifiedshift}, this plot shows how fast initial CMC and \statio data get to their final state. {Apart from the two solid lines representing (data also appearing in \sfref{nsH}), two simulations are included: CMC initial data, and either i) (dashed line) a different value of parameter $\kappa_1=1.5$ (taken to be 0.5 otherwise), or ii) (dash-dot line) evolved slicing condition with parameter values $m_{cK}=0.1$ and $\cnK=0.5$}. \textbf{(B)} Values of the Hamiltonian constraint over time at different values of the compactified radial coordinate: in the middle of the domain for dash-dot, near \scrip for dashed and at the closest point to \scrip for the solid line. Constraint violations are larger closer to future null infinity.}
\label{ns}
\end{figure}

Both sets of initial data are now evolved with the shift condition \eref{modifiedshift}. The initial dynamics in the CMC case are driven by the slicing condition \eref{adotsolve}, while the \statio solution remains visually static throughout the evolution (ran up to $t=1000$). 
Figure \ref{nsL2} aims to capture how fast the final state is attained in the evolutions, via showing the behaviour of the $L^2$ norm over the whole $\Lambda^r$ gridfunction over time, for both sets of initial data {(CMC and \statio), as well as for two runs with CMC initial data and different parameter configurations}. The \statio $\Lambda^r$ {is set to be exactly zero initially (implied by the explicit conformal flatness imposed), although this cannot be seen in \fref{nsL2} due to the logarithmic plot used in the veritcal axis. As the evolution starts its value changes,} probably because the solved-for solution has some small errors and still needs to relax to its truly stationary state. In any case, the change is much smaller than for CMC initial data. Both sets of initial data relax to the final state at the same rate. The quality of the final solution is estimated looking at the values of the Hamiltonian constraint at different values of the radial coordinate as time passes in \fref{nsH}. In the middle of the compactified domain (dash-dot line), the value of $H$ rapidly approaches a small value. Closer to \scrip (dashed line), the final value is larger, meaning the constraint violation there is more pronounced than in the centre (which is expected, because the equations are formally singular). At the last gridpoint (half a spatial step from the compactified location of future null infinity, solid line), the effect is even larger. One indication that the \statio solution must have some small errors is that in the blue solid line the leftmost value of $H$ for CMC data (at $t=0$) is very small -- except for the compactification factor, which is solved numerically, the rest is an explicit analytic solution \eref{varshypcompc}. However, the initial value of the Hamiltonian constraint for \statio data on the black solid line is large, indicating that the given initial data does not satisfy the constraints in a satisfactory way, at least very near \scrip. 

\begin{figure}[h!]
\begin{center}
\includegraphics[width=0.9\textwidth]{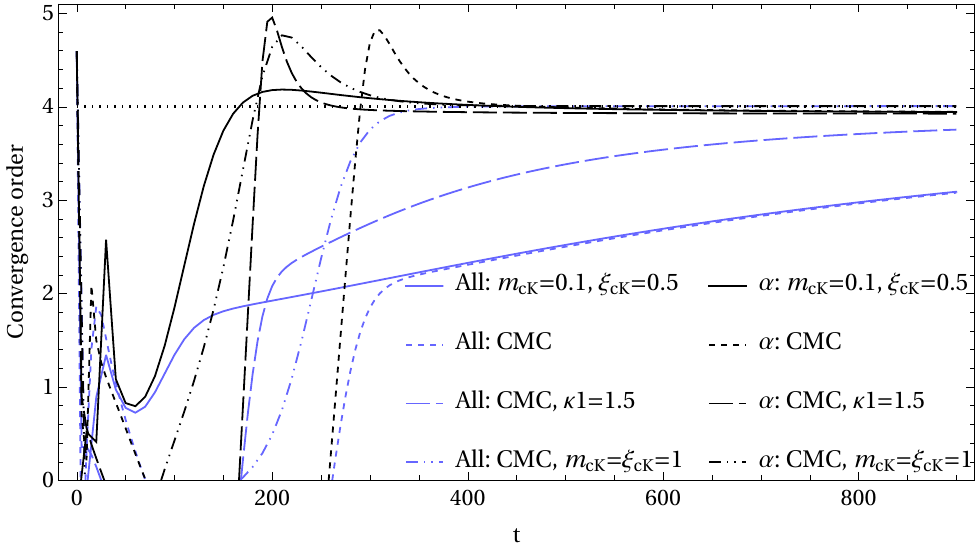}
\end{center}
\caption{{Convergence order over time calculated using the $L^2$-type norm as $\log_f\left(\frac{\sqrt{\sum_{i=1}^N(X_{low, i}-X_{med, i})^2}}{\sqrt{\sum_{i=1}^N(X_{med, i}-X_{high, i})^2}}\right)$ with $i\in[1,N]$, $f=1.5$, $N=304$ and $X_{low/med/high, i}$ denoting the evolution variables for the low, medium and high resolutions at point $i$. For the blue curves, all the evolution variables were included (so that $X$ was successively $\chi,\gamma_{rr},A_{rr}\K,\Lambda^r,\alpha$ and $\beta^r$), while for the black lines the summation was performed only on $\alpha$. 
The expected 4th order convergence is lost very early, and only recovered later in the evolution, if at all. Interpretation in the main text. 
Further understanding the convergence behaviour in the evolution for the choice of gauge conditions is beyond the scope of this work.}}\label{norm}
\end{figure}

{The dashed and dash-dot lines included in \sfref{nsL2} correspond to two simulations with CMC initial data, but with a different parameter configuration: respectively $\kappa_1=1.5$, and slicing with $m_{cK}=\cnK=1$. They have been included to shed light on the loss of self-convergence issues detected in the simulations, and shown in \fref{norm}. The $L^2$-type norm of the stationary states of $\Lambda^r$ for these two different configurations are different from the solid lines (that for the slicing with $m_{cK}=\cnK=1$ gets closer to zero).}

{The convergence order calculated from the evolution variables shown in \fref{norm} drops from the expected 4 very early in the simulation. Self-convergence of the lapse (black lines) is recovered around $t\sim250$. Other fields behave worse (not shown here): for instance and except for the CMC $m_{cK}=\cnK=1$ runs, self-convergence of $\gamma_{rr}$ is only 2nd order, while $A_{rr}$'s is between 2nd and 4th. This is why the convergence order that takes into account all of the eolution variables does not go back to 4 even by $t=900$. The exception is the CMC $m_{cK}=\cnK=1$ case, where 4th order convergence is recovered around $t\sim300$. The reason is that a larger value of $\cnK$ damps the deviations of the lapse from its value at \scrip more efficiently, together with a larger propagation speed near the trumpet given by a bigger $m_{cK}$. With just the simple slicing condition \eref{adotsolve} considered here one can appreciate the enormous effect that gauge conditions have on the evolutions. 
Still, convergence is lost before $t\sim300$ even in the best case, which is clearly pointing to the presence of a problem. To the author's best knowledge, the shift condition \eref{modifiedshift} has not been used before, and could be partially to blame of the loss of convergence. Other more sophisticated gauge conditions, such as those covered in \cite{\papgauge}, could be tested for comparison -- although the \statio solution would not be a stationary solution anymore. 
During the time when self-convergence of the variables was lost virtually everywhere in the domain ($0\lesssim t\lesssim 300$ or later), the constraints continued to converge in the interior at the domain at all times. However, at late times they did not asymptote to zero, but to $\sim10^{-6}$. Understanding all these interesting aspects and how they relate to the choice of gauge conditions is left for future research.}

To study the robustness of the gauge system and study how {constraint} convergence evolves in the simulations, a constraint-satisfying Gaussian-like gauge perturbation is included in the initial lapse, 
\begin{equation}\label{e:inipertutb}
\alpha= \alpha_0 + A e^{-\frac{(r^2-r_c^2)^2}{4\sigma^4}}, 
\end{equation}
where here $\alpha_0$ denotes either its CMC value as in \eref{varshypcompc} or its \statio value. The chosen values of the parameters are $A=0.05$, $\sigma=0.1$ and $r_c=0.25$. This initial data is run with 304, 456 and 684 spatial points (and corresponding timestep $dt = 10^{-3}, 6.667\cdot 10^{-4}, 4.444\cdot 10^{-4}$), so increasing the resolution by $1.5$ between runs. 

\begin{figure}
    \begin{subfigure}[t]{0.49\textwidth}
        \includegraphics[width=\linewidth]{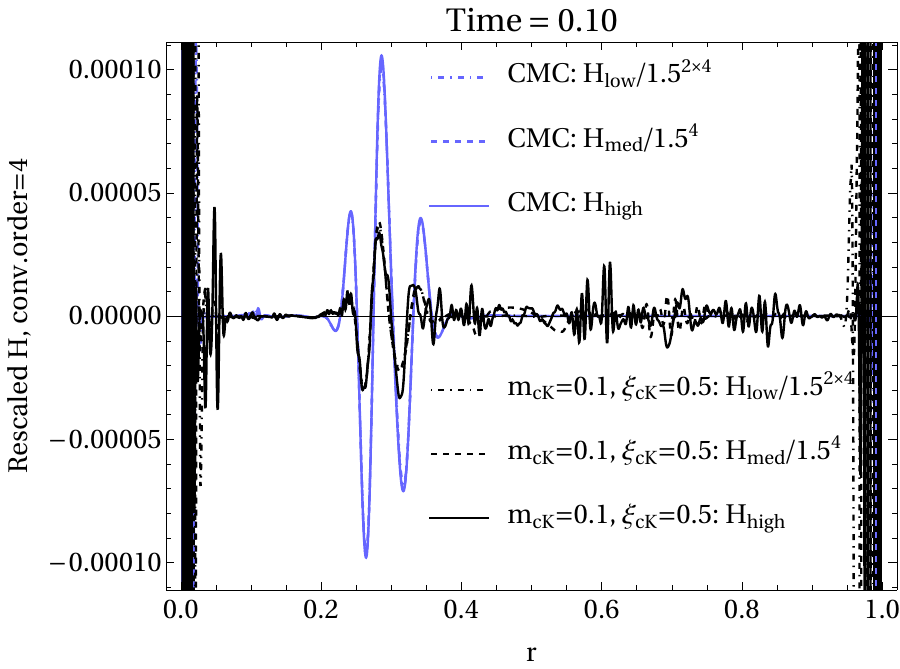}
        \caption{Rescaled H constraint at $t=0.1$.}
        \label{Hgauge1}
    \end{subfigure}  
    \begin{subfigure}[t]{0.49\textwidth}
        \includegraphics[width=\linewidth]{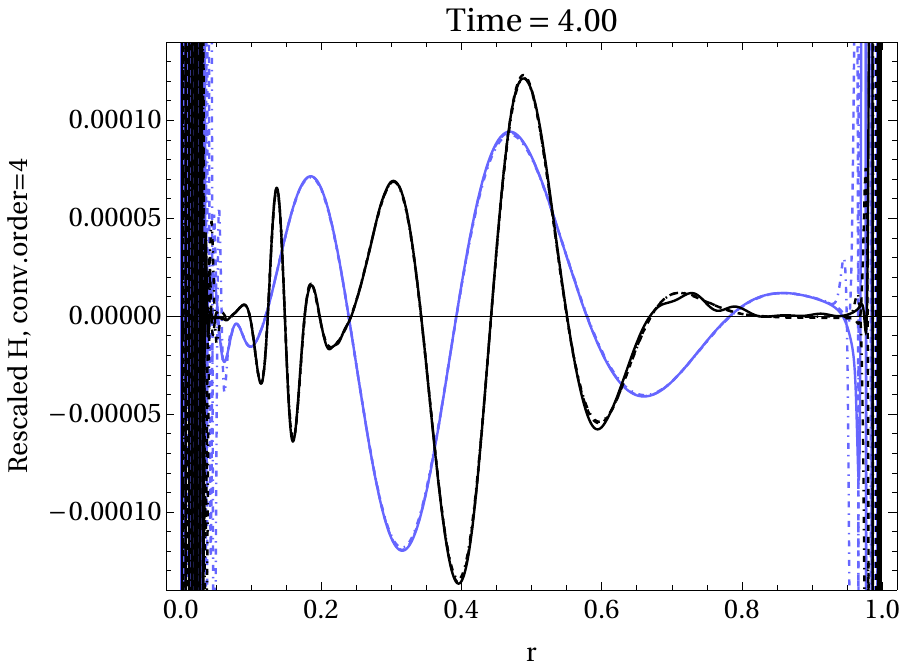}
        \caption{Rescaled H constraint at $t=4$.}
        \label{Hgauge2}
    \end{subfigure}
\caption{Rescaled values of the Hamiltonian constraint as a function of $r$ at two different instants of time. $H_{low}$ uses 304 points, $H_{med}$ uses 456, and $H_{high}$ uses 684.  The rescalings are of the form $f^p$, where $f$ is the increase in resolution between runs ($1.5$ in this case) and $p$ is the order of convergence (4th in this setup). The same legend is valid for both plots.}
\label{Hgauge}
\end{figure}

The initial gauge perturbation extends to all evolution variables and gets propagated away, part into the BH and part out through \scrip, leaving behind what under visual inspection looks like the \statio solution. Even if the \statio initial data do not appropriately converge, evolution of the gauge perturbation will after a certain amount of time. Looking at convergence of the constraints can give an estimate of when that happens, as well as the general reliability of the simulation. Figure \ref{Hgauge} shows convergence of the Hamiltonian constraint at an early time $t=0.1$ and a later one $t=4$, for both CMC and \statio initial data. 
Except at the boundaries, convergence for the CMC case is good, as the blue curves coincide very well in the interior of the domain for both figures. That is not the case for the \statio case: initially the lack of convergence of the initial data is seen clearly in the non-coincidence between the black curves in \fref{Hgauge1}. Later, at $t=4$ as shown in \fref{Hgauge2}, most of the discrepancies have disappeared and convergence looks better -- except again at the extrema of the radial coordinate, where the Hamiltonian constraint looks very noisy. This is not necessarily an indicator of a problem, as the {constraints} are formally divergent at the trumpet and at \scrip\footnote{{Constraint equations, zero for the continuum solution, do not have a scale. Self-convergence of the evolution fields is what needs to be satisfactory there. As described above, this is not the case, but this will be studied elsewhere.}}. 

\section{Conclusions}\label{conclusions}

The hyperboloidal approach allows numerical simulations to reach future null infinity from first principles and without complicated constructions. While it needs to be further understood and developed, progress in the non-linear regime is taking place along several fronts \cite{Peterson:2023bha,Feng:2023ppe,3dpaper}, and will also benefit from work in the linear one, like e.g.~\cite{Jaramillo:2020tuu}.
The focus in the present work has been on spherically symmetric hyperboloidal trumpet initial data for puncture-type evolutions of BHs. 
More specifically, the construction of CMC trumpet data via the height function approach has been reviewed and adapted to the needs of numerical simulations using the puncture approach together with conformal compactification. 
Gauge conditions play a crucial role in numerical evolutions, both in the stability of the setup and the allowed final states of the system. While understanding them within the hyperboloidal approach is still work in progress, some options providing successful numerical simulations are known. 

Availability of stationary initial data that are suitable for the evolution and study of perturbations thereof is very desired, especially as part of the development of the hyperboloidal method. This work sets the infrastructure to pursue those solutions for spherically symmetric BH trumpet data within the conformally compactified domain. 
A procedure to solve a specific numerically-tested slicing condition together with explicit conformal flatness has been developed and tested in examples. The accuracy of the numerical solutions was not fully satisfactory {(convergence could only be checked in part of the domain)}, but still they could be tested in hyperboloidal evolutions and shed some light on the behaviour of some choices of shift conditions. 
{The bottomline is that new initial data that approaches the stationary solution of the gauge conditions much more rapidly than CMC data was constructed, despite how challenging the procedure ultimately was.}

There are several options that could be tested to improve the quality of the stationary solutions and that have been left for future work. At the numerical level, a more sophisticated solving method could be used, such as an elliptic solver for non-linear equations or a relaxation method. The main requirement is to obtain reliable {and smooth} solutions for which convergence can be checked {in the whole integration domain}. At the analytical level, a different slicing condition could be considered. To be suitable, it needs to provide a stable stationary final state in evolution for a hyperboloidal trumpet slice. This is not easy to attain, but progress on the gauge condition front in the near future will provide more options. {This progress will also contribute to understand the convergence problems detected during the evolutions, which will be solved elsewhere.}

The insight gained on the effects of the two different shift conditions tested here will be used to develop other suitable options that may provide smoother profiles of the evolution variables near \scrip. Within the free evolution setup with a BSSN/Z4-type formulation, the shift condition is closely related to the quantity $\Lambda^a$, which may impose some limitations as to which final states are allowed or not. Either modifying the definition of $\Lambda^a$, making it more compatible with the hyperboloidal framework, or considering a different formulation of the Einstein equations may be beneficial. In either case, it would be worth attempting to solve the shift condition for the compactification, once the behaviour of $\Lambda^a$ is better understood, instead of imposing conformal flatness as done here. 

While CMC trumpet initial data for the RN spacetime has been developed, it is still waiting to be tested in hyperboloidal evolution, to the author's best knowledge. Also, how far the conformally compactified height function approach can be extended to include a non-vanishing cosmological constant, in a similar way to e.g. \cite{Bizon:2020qnd} but with views towards non-linear numerical simulations, is still to be found out. 
Finally, including a massless scalar field perturbation on stationary trumpet initial data will allow to study the former's behaviour without mixing it with trumpet relaxation dynamics, which was one of the problems hit in \cite{\procere,\thesis}. Of special relevance are the decay tails and their convergence at \scrip, as preparation of the GWs to be treated in the full 3D case. 

{This work succeeded in taking a few steps towards a more thorough understanding of the interplay between gauge conditions and stationary solutions on hyperboloidal trumpet slices for the puncture approach. It has provided a framework to determine those solutions for suitable generic gauge conditions, as well as exemplifying how challenging both the initial data calculations and the evolutions are. Insights into how to develop better suited formulations for the hyperboloidal approach have also been gained.}


\section*{Acknowledgments}

The author thanks Edgar Gasperin, Sascha Husa and David Hilditch for valuable comments on the manuscript. 
The author would also like to thank Sascha Husa for important feedback in parts of the research presented here. Discussions with with him, Sergio Dain, Niall O’Murchadha and David Hilditch considerably helped understand and solve challenges in this work. 
Most of the algebraic derivations were performed using the Mathematica package \texttt{xAct} \cite{xAct}.
The author thanks the Fundac\~ao para a  Ci\^encia e Tecnologia (FCT), Portugal, for the financial support to the Center for Astrophysics and Gravitation (CENTRA/IST/ULisboa) through the Grant Project~No.~UIDB/00099/2020. 
This work was also supported through the European Research Council Consolidator Grant 647839. 

\appendix

\section{Initial data equation from slicing condition}\label{eqap}

Equation resulting from substituting relations \eref{substatio} into \eref{adotsolve} and solving for $\beta'$:  
\begin{eqnarray}\label{eqlong}
\beta' =&& \frac{1}{9 \alpha \,c\,r \Omega ^2 \bar\omega^2 \left(c^2 \Kc \Omega  (2 M \Omega  \bar\omega-1)+6 m_{cK}   \rscri\right)}
\left[\alpha \,c\,\Kc \bar\omega \left(-9 \Omega ^2 \left(c^2 (\alpha  \Kc-\cnK)   \right.\right.\right.\nonumber \\ && \left.\left.\left.
+\frac{3 \alpha  \cnK}{\sqrt{\Kc^2 r^2+9 \Omega ^2}}\right)-18 c^2 M \Omega ^3 \bar\omega (\cnK-\alpha  \Kc)-\frac{\alpha    \Kc^2 r^2 \left(3 \cnK+\Kc r \Omega '\right)}{\sqrt{\Kc^2 r^2+9 \Omega ^2}}    \right.\right.\nonumber \\ && \left.\left.
+\Omega  \left(\frac{\alpha  \Kc^3 r^2}{\sqrt{\Kc^2 r^2+9 \Omega ^2}}+54 \alpha  m_{cK} \rscri\right)\right)-9 \beta ^3 \Omega  \bar\omega^2 \left(c^2   \Kc \Omega  (9 M \Omega  \bar\omega-4)+12 m_{cK} \rscri\right)   \right.\nonumber \\ && \left.
+9 \beta \,c\,\Omega  \bar\omega^2 \left(10 c^3 \Kc M^2 \Omega ^5 \bar\omega^2+2 \Omega ^3 \left(c^3 \Kc+12\,c\,M m_{cK} \rscri   \bar\omega\right)-9 c^3 \Kc M \Omega ^4 \bar\omega   \right.\right.\nonumber \\ && \left.\left.
-\alpha  c^2 \Kc r \Omega  \Omega '+2\,c\,\Omega ^2 \left(\alpha \,c\,\Kc M r \bar\omega \Omega '-6 m_{cK} \rscri\right)+6 \alpha  m_{cK} r \rscri \Omega   '\right)      \right.\nonumber \\ && \left. 
+9 \alpha  \beta ^2\,c\,\Kc \bar\omega (\cnK-\alpha  \Kc)+18 \beta ^5 \Kc \bar\omega^2\right] . 
\end{eqnarray}
Here $\alpha$ was kept for readability and is to be substituted using \eref{substatio}. The coordinate location of \scrip at $\rscri=1$ is also to be set.

\bibliography{../../hypcomp.bib}

\end{document}